\documentclass[12pt]{article}
\usepackage{color}
\usepackage{latexsym}
\usepackage{epsfig,amssymb,euscript, mathrsfs}
\usepackage{amsmath}
\usepackage{bbm}
\newsavebox{\ns}
\newsavebox{\dbrane}
\newsavebox{\dbshort}

\def\be{\begin{equation}}
\def\ee{\end{equation}}
\def\bea{\begin{eqnarray}}
\def\eea{\end{eqnarray}}

\newcommand{\nn}{\nonumber}

\newcommand\R{\mathbb{R}}
\newcommand\Z{\mathbb{Z}}

\newcommand\C{\mathbb{C}}

\newcommand\diff{\mathrm{d}}
\newcommand{\de}{\partial}

\newcommand{\dd}{\mathrm{d}}
\newcommand{\me}{\mathrm{e}}
\newcommand{\ii}{\mathrm{i}}

\newcommand{\ex}{\mathrm{e}}
\newcommand{\vol}{\mathrm{vol}}

\newlength{\sswidth}

\newcommand{\Db}{\Delta_{\mathrm{bos}}}
\newcommand{\Df}{\Delta_{\mathrm{fer}}}
\newcommand{\hL}{\hat{\mathcal{L}}}
\newcommand{\bY}{\overline{Y}}
\newcommand{\bK}{\overline{K}}

\newcommand{\Bold}{\check B}
\newcommand{\phiold}{\check\phi}
\newcommand{\Phiold}{\check\Phi}

\newcommand{\Esusy}{E_\mathrm{susy}}
\newcommand{\iflat}{\, \alpha}
\newcommand{\Mflat}{\mathcal{M}_{\mathrm{flat}}}
\newcommand{\holonomy}{\nu}
\newcommand{\Morb}{\mathbb{M}}
\newcommand{\Nlambda}{q}
\newcommand{\anothers}{\sigma}
\newcommand{\Esusym}{E_\mathrm{susy}^{\mathrm{matter}}}
\newcommand{\Esusymm}{E_\mathrm{susy, \, \mathbf{m}}^{\mathrm{matter}}}

\newcommand{\Lef}{C}

\newcommand{\qq}{k}
\newcommand{\Deltaqu}{\Xi}

\numberwithin{equation}{section}       

\begin{document}

\begin{titlepage}

\begin{center}

\today

\vskip 2.3 cm 

\vskip 5mm

{\Large \bf The character of the }

\vskip 5mm

{\Large \bf supersymmetric Casimir energy}

\vskip 15mm

{Dario Martelli${}^{\,a}$ and James Sparks${}^{\,b}$}

\vskip 1cm

${}^a$\textit{Department of Mathematics, King's College London, \\
The Strand, London, WC2R 2LS,  UK\\}

\vskip 1cm

${}^{\,b}$\textit{Mathematical Institute, University of Oxford,\\
Andrew Wiles Building, Radcliffe Observatory Quarter,\\
Woodstock Road, Oxford, OX2 6GG, UK\\}

\end{center}

\vskip 1 cm

\begin{abstract}
\noindent  We study the supersymmetric Casimir energy $\Esusy$ of $\mathcal{N}=1$ field theories with an R-symmetry, defined on rigid supersymmetric backgrounds 
$S^1\times M_3$, using  a Hamiltonian formalism. These backgrounds admit an ambi-Hermitian geometry, and we  show that the net contributions to $\Esusy$ arise from certain
twisted holomorphic modes on $\R\times M_3$, with respect to both complex structures. The supersymmetric Casimir energy may then be identified
as a limit of an index-character that counts these modes. In particular this explains a recent observation relating $\Esusy$ 
on $S^1\times S^3$ to the anomaly polynomial. As further applications we compute $\Esusy$ for certain secondary Hopf surfaces, and  discuss
 how the index-character may also be used to compute  generalized supersymmetric indices.

\end{abstract}

\end{titlepage}

\pagestyle{plain}
\setcounter{page}{1}
\newcounter{bean}
\baselineskip18pt
\tableofcontents

\newpage

\section{Introduction}

In recent years the technique of localization \cite{Pestun:2007rz} has provided access to a host of exact results in supersymmetric field theories defined on certain curved backgrounds. 
This method can be used to compute  a number of observables in strongly coupled field theories. 
These in general depend on the background geometry, leading to a richer structure than in  flat space. 
In this paper we will consider the supersymmetric Casimir energy, introduced in \cite{Assel:2014paa} and further studied in  \cite{Lorenzen:2014pna,Assel:2015nca,Bobev:2015kza}.
We will focus on four-dimensional $\mathcal{N}=1$ theories with an R-symmetry, defined 
on manifolds $S^1\times M_3$,  with $M_3$ a compact three-manifold. These arise as rigid supersymmetric backgrounds
admitting two supercharges of opposite R-charge, which  are ambi-Hermitian, with integrable complex 
stuctures $I_\pm$    \cite{Klare:2012gn,Dumitrescu:2012ha}. Moreover, the backgrounds are  equipped with a complex Killing vector field $K$ of Hodge type $(0,1)$ for 
both complex structures.  Denoting this as  $K=\frac{1}{2}(\xi-\ii \partial_\tau)$, where $\tau\in[0,\beta)$ parametrizes $S^1=S^1_\beta$, 
$\xi$ is a nowhere zero vector on $M_3$ (the \emph{Reeb} vector field), generating a transversely holomorphic foliation. 
When all orbits of $\xi$ close, this means that $M_3$ is a Seifert fibred three-manifold, with $\xi$ generating the fibration. 

On such a background,  one can  consider the partition function of an $\mathcal{N}=1$  theory with supersymmetric boundary 
conditions for the fermions. As is familiar from finite temperature field theory, this computes
\bea\label{Zintro}
 Z_{S^1_\beta \times M_3} & = & \mathrm{Tr}\, \ex^{-\beta H_{\mathrm{susy}}}~,
\eea
where the Hamiltonian $H_{\mathrm{susy}}$ generates time-translations along $\partial_\tau$. Supposing this 
has a spectrum of energies $\{E_i\}_{i\in I}$, with $H_{\mathrm{susy}}\, |i\rangle = E_i \, |i  \rangle$, then the minimum energy is $E_0\equiv E_{\mathrm{susy}}$ where evidently
\bea
E_{\mathrm{susy}} &=   & -\lim_{\beta\rightarrow\infty} \frac{\diff}{\diff\beta} Z_{S^1_\beta\times M_3}.
\eea
Thus the supersymmetric Casimir energy  is given by $E_{\mathrm{susy}} = \langle\,  0| H_{\mathrm{susy}}|0\rangle$, where $|0\rangle$ is the vacuum state. 
Unlike the usual Casimir energy on $S^1\times M_3$ (proportional to the  integral of the energy-momentum tensor $T_{\tau\tau}$ over $M_3$), this has been argued to be a well-defined observable of the
theory, {\it i.e.} it is scheme-independent, in any supersymmetric regularization \cite{Assel:2015nca}. 

We will be interested in computing $\langle H_{\mathrm{susy}} \rangle = E_{\mathrm{susy}}$ 
via canonical quantization. This approach was initiated in \cite{Lorenzen:2014pna} for the conformally flat $S^1\times S^3$ background, and further 
elaborated on in  \cite{Assel:2015nca}. 
One can dimensionally reduce the one-loop operators on $M_3$ to obtain a  supersymmetric quantum mechanics 
on $\R_{\tau}$, where the $\beta\rightarrow\infty$ limit effectively decompactifies the circle $S^1_\beta$. 
Most of the modes of the one-loop operators are paired by supersymmetry, 
and these combine into long multiplets that do not contribute to $\langle H_{\mathrm{susy}} \rangle$ in the 
supersymmetric quantum mechanics \cite{Assel:2015nca}. In this paper we will show that the unpaired modes are certain (twisted)
holomorphic functions on $\R\times M_3$, where there is one set of modes for each of the two complex 
structures $I_\pm$. More precisely, here will restrict attention to the contribution of the chiral multiplet. We expect that the vector multiplet contributions will also arrange into short multiplets, and will similarly be  related to 
(twisted) holomorphic functions. However, we will not perform this analysis in this paper. 

When $\R\times M_3\cong X\setminus\{o\}$ is the complement of an isolated singularity $o$ in a Gorenstein canonical 
singularity $X$, one can elegantly solve for these unpaired modes that contribute to the supersymmetric Casimir energy.  
These include of course $M_3= S^3$, as well as  $M_3=L(p,1)=S^3/\Z_p$ (\emph{i.e.} a Lens space), 
for which $X=\C^2$ and $X=\C^2/\Z_p$ is an $A_{p-1}$ singularity, 
previously studied in the literature; but this construction also includes  many other interesting three-manifolds.
 A large class may be constructed from homogeneous hypersurface singularities. 
Here $X$ comes equipped with a $\C^*$ action, which is generated by the complex vector field $K$, and $X\setminus\{o\}$ fibres 
over a compact orbifold Riemann surface $\Sigma_2$. Then $X_+\cong X_-\cong X$ are isomorphic as complex varieties, 
but the  relative sign of the complex structures on fibre and base are opposite in the two complex structures $I_\pm$.
We will show that the modes that contribute to the supersymmetric Casimir energy 
in a chiral matter multiplet take the form
\bea\label{Phipmmodes}
\Phi_\pm & = & \left|\frac{\mathcal{P}_\pm}{\Omega_\pm}\right|^{\pm \qq_\pm/2} \mathscr{F}_\pm~,
\eea
where $\mathcal{P}_\pm$ are the globally defined nowhere zero $(2,0)$-forms defined by the Hermitian 
structures for $I_\pm$, while $\Omega_\pm$ are the globally defined nowhere zero holomorphic $(2,0)$-forms 
of definite Reeb weight under $\xi$, that exist because $X_+\cong X_-$ is Gorenstein. 
Furthermore, $\qq_\pm $ denote the R-charges of the relevant fields; in particular, $\qq_+=r-2$, $\qq_-=r$, where $r\in \R$ is the R-charge 
of the top component of a  chiral multiplet. These correspond to fermionic ($\Phi_+$) and bosonic ($\Phi_-$) modes, respectively. 
The essential point in (\ref{Phipmmodes}) is that $\mathscr{F}_\pm$ are simply \emph{holomorphic functions} on $X_\pm$. 
More precisely, in general the path integral (\ref{Zintro}) 
 splits into different topological sectors, labelled by flat gauge connections, and 
 for the trivial flat connection $\mathscr{F}_\pm$ are holomorphic functions; more 
 generally they are holomorphic  sections of the associated flat holomorphic bundles. 
 For example, for quotients of $M_3\cong S^3$, such as the Lens spaces $L(p,1)=S^3/\Z_p$, 
the relevant holomorphic modes may be obtained as a 
 projection of the holomorphic functions on the covering space.

The supersymmetric Casimir energy is computed by ``counting'' these holomorphic functions 
according to their charge under the Reeb vector $\xi$. As such, $E_{\mathrm{susy}}$ is 
closely related to the index-character of \cite{Martelli:2006yb}. In this reference, 
it was shown that the volume  of a Sasakian  manifold $Y$ can be obtained from a certain 
 limit  of the equivariant index of the $\bar \de$ operator on the associated K\"ahler cone singularity $X=C(Y)$. 
In a similar vein, here we will show that the  supersymmetric Casimir energy is obtained from a limit of an 
index-character counting holomorphic functions on $\R\times M_3$. In the case of $M_3\cong S^3$, 
this  explains  a conjecture/observation  made in   \cite{Bobev:2015kza}, where it was proposed that  $E_{\rm susy}$ 
may be computed using the equivariant anomaly polynomial. 

 The rest of the paper is organized as follows. 
In section \ref{sec:SUSYbackground} we review and expand on the relevant background geometry, emphasizing the role of the ambi-Hermitian structure.  
In section \ref{sec:Casimir}, after recalling how the supersymmetric Casimir energy arises, we formulate the conditions for (un-)pairing of modes 
on $\R\times M_3$. In section \ref{sec:primary} we discuss the index-character counting holomorphic functions, and make the connection with 
  \cite{Bobev:2015kza} in the case of primary Hopf surfaces.  Extensions to secondary Hopf surfaces, and more general $M_3$ realized as links of homogeneous 
  hypersurface singularities, are discussed in section~\ref{moregeneral:sec}.
We conclude in section \ref{discuss:sec}. We have included an Appendix  \ref{letters:sec}, where we discuss the relation of the index-character to the supersymmetric index \cite{Romelsberger:2005eg} and its generalizations.

\section{Supersymmetric backgrounds}\label{sec:SUSYbackground}

\subsection{Background geometry}\label{sec:background}

We are interested in studying four-dimensional $\mathcal{N}=1$ theories with an R-symmetry on 
 $M_4=S^1\times M_3$, where $M_3$ is a compact three-manifold.  
 In Euclidean signature, the  relevant supersymmetry conditions  are the two independent first-order differential equations
\bea
( \nabla_{\mu} \mp \ii A_{\mu} ) \zeta_\pm  + \ii V_{\mu} \zeta_\pm + \ii V^{\nu} (\sigma_\pm)_{\mu\nu} \zeta_\pm \!& = &\! 0  \ , \label{KeqnZeta}
\eea
where $\zeta_\pm$ are spinors of opposite chirality. Here we use the spinor conventions\footnote{Differently from previous literature, 
we denote the Killing spinors and associated  complex structures with $\pm$ subscripts. This emphasizes the fact that the two spinors and complex structures 
are on an equal footing.} 
of \cite{Dumitrescu:2012ha}, 
in which $\zeta_\pm$ are two-component spinors 
with corresponding Clifford algebra generated by $({\sigma_\pm})^a=(\pm\vec{\sigma},-\ii \mathbbm{1}_2)$, 
where $a=1,\ldots,4$ is an orthonormal frame index and $\vec{\sigma}=(\sigma^1,\sigma^2,\sigma^3)$ are 
the Pauli matrices. In particular the generators of $SU(2)_\pm\subset \mathrm{Spin}(4)=SU(2)_+\times SU(2)_-$ are 
\bea
(\sigma_\pm)^{ab} &=& \frac{1}{4}\left(\sigma_\pm^a\sigma_\mp^b - \sigma_\pm^b\sigma_\mp^a\right)~.
\eea 
The field $V_\mu$ is assumed to be a  globally defined  one-form obeying $\nabla^{\mu} V_{\mu} =0$, and will not play a role in this paper. 
The field  $A_\mu$ is associated to local R-symmetry transformations, with all matter fields being charged under this via appropriate covariant derivatives. 

The Killing spinors $\zeta_\pm$  equip $M_4$ with  two commuting integrable complex structures\footnote{We adopt the same sign conventions as \cite{Closset:2013sxa, Closset:2013vra} 
for the complex structures. Our main motivation for this choice of convention is that the modes that contribute to the supersymmetric Casimir energy will
turn out to be (twisted) \emph{holomorphic}, whereas if we reversed the signs of the complex structures they would be (twisted) anti-holomorphic.\label{fucsigns}}
\bea\label{Ipm}
(I_\pm)^\mu_{\ \, \nu} &\equiv & -\frac{2\ii}{|\zeta_\pm|^2}\zeta_\pm^\dagger(\sigma_\pm)^\mu_{\ \, \nu}\zeta_\pm~.
\eea 
The metric $g_{M_4}$ is Hermitian with respect to both $I_\pm$, 
but where the induced orientations are opposite, which means the geometry is by definition \emph{ambi-Hermitian}. 
 This structure also equips $M_4$ 
with  a complex Killing vector field 
\bea\label{K}
K^\mu & \equiv & \zeta_+\sigma_+^\mu\zeta_-~.
\eea
This has Hodge type $(0,1)$ for 
both complex structures, and satisfies  $K^\mu K_\mu=0$. We assume that $K$ commutes with its complex conjugate $K^*$, $[K,K^*]=0$.\footnote{If $[K,K^*]\neq 0$ the 
metric is locally isometric to $\R\times S^3$ with the standard round metric on $S^3$ \cite{Dumitrescu:2012ha}.}
It then follows that we may write
 $K=\frac{1}{2}(\xi-\ii \partial_\tau)$, where $\tau\in[0,\beta)$ parametrizes $S^1=S^1_\beta$
and $\xi$ is a nowhere zero vector field on $M_3$. 

Following \cite{Assel:2014paa}, we assume the metric 
 on $M_4=S^1\times M_3$ to take the form
\bea\label{4dmetric}
g_{M_4} &=& \Omega^2\left(\diff \tau^2 + g_{M_3}\right)~,
\eea
where the local form of the metric on $M_3$ may be written as
\bea\label{3dmetric}
g_{M_3} &=& (\diff \psi+a)^2 + c^2\diff z\diff\bar{z}~.
\eea
Here $\xi=\partial_\psi$ generates a transversely holomorphic foliation of $M_3$, with 
 $z$ a local transverse complex coordinate. Since $\partial_\tau$ and 
$\partial_\psi$ are both Killing vectors the positive conformal  factor is $\Omega=\Omega(z,\bar{z})$,
 while $c=c(z,\bar{z})$ is a locally defined non-negative function and $a=a_z(z,\bar{z})\diff z+ \bar{a}_{\bar{z}}(z,\bar{z})\diff\bar{z}$ is 
a local real one-form.
Notice that any Riemmanian three-manifold admitting a unit length Killing vector $\xi=\partial_\psi$ may be put into the
local form (\ref{3dmetric}). Notice also that this geometry is precisely the rigid three-dimensional supersymmetric 
geometry of \cite{Closset:2012ru, Alday:2013lba}, for which  there are two three-dimensional supercharges of opposite R-charge.

We shall refer to $\xi=\partial_\psi$ as the \emph{Reeb} vector field. Globally the foliation of $M_3$
that it induces splits into three types: regular, quasi-regular and  irregular. In the first two cases 
all the leaves are closed, and hence $\xi$ generates a $U(1)$ isometry of $M_3$. If this $U(1)$ action is 
free, the foliation is said to be regular. In this case $M_3$ is the total space of a circle bundle over a 
compact Riemann surface $\Sigma_2$, which can have arbitrary genus $g\geq 0$. The local metric
$c^2\diff z\diff\bar{z}$ then pushes down to a (arbitrary) Riemannian metric on $\Sigma_2$, while 
the one-form $a$ is a connection for the circle bundle over $\Sigma_2$. More generally, in the quasi-regular case
since $\xi$ is nowhere zero the $U(1)$ action on $M_3$ is  necessarily  locally free, 
and the base $\Sigma_2\equiv M_3/U(1)$ is an orbifold Riemann surface. 
Topologically this is a Riemann surface of genus $g$, with some number $\Morb$ of orbifold points 
which are locally modelled on $\C/\Z_{k_i}$,  $k_i\in\mathbb{N}$, $i=1,\ldots,\Morb$. The induced metric on $\Sigma_2$ then has a conical deficit around each orbifold point, with 
total angle $2\pi/k_i$. The three-manifold $M_3$ is  the total 
space of a circle orbibundle over $\Sigma_2$. Such three-manifolds are called \emph{Seifert fibred three-manifolds}, 
and they are classified. 

In the irregular case $\xi$ has at least 
one open orbit. Since the isometry group of a compact manifold is compact, this means that 
$M_3$ must have at least $U(1)\times U(1)$ isometry, with $\xi$ being an irrational linear combination 
of the two generating vector fields. Notice that $M_3$ is still a Seifert manifold, by taking a rational 
linear combination, and that the corresponding base $\Sigma_2$  inherits a $U(1)$ isometry. 
There are then two cases: either this $U(1)$ action is Hamiltonian, meaning there is an associated moment map, 
or else $\pi_1(\Sigma_2)$ is non-trivial. In the first case $\Sigma_2\cong \mathbb{WCP}^2_{[p,q]}$ is 
necessarily a weighted projective space \cite{LT}, while in the second case instead $\Sigma_2\cong T^2$. 
In particular in the first case $M_3$ is either $S^1\times S^2$, or it has finite fundamental group with 
simply-connected covering space $S^3$.

In addition to the local complex coordinate $z$, we may also introduce
\bea
w & \equiv & \psi - \ii \tau + P(z,\bar{z})~,
\eea
where $P(z,\bar{z})$ is a local complex function. Taking this to solve
\bea
\partial_z\overline{P} &=& a_z~,
\eea
where recall that $a$ is the local one-form appearing in the metric (\ref{3dmetric}), and defining
\bea
h &=& h(z,\bar{z}) \ \equiv \ -2\ii \, \partial_z \mathrm{Im}\, P~,
\eea
the metric (\ref{4dmetric}) may be rewritten as
\bea
g_{M_4} &=& \Omega^2\left[(\diff w + h \diff z)(\diff \bar{w}+\bar{h} \diff\bar{z}) + c^2\diff z\diff\bar{z}\right]~.
\eea
In these complex coordinates we have the complex vector fields
\bea
K &=& \partial_{\bar{w}}~, \qquad Y \ = \ \frac{s}{\Omega^2 c}(\partial_{\bar{z}} -\bar{h}\partial_{\bar{w}})~.
\eea
Here $s$ is a complex-valued function which appears in the Killing spinors $\zeta_\pm$, 
where  the vector $Y$, like $K$ in (\ref{K}), is defined as a spinor bilinear via
\bea
Y^\mu &\equiv & \frac{1}{2|\zeta_-|^2}\zeta_-^\dagger \sigma_-^\mu \zeta_+~.
\eea
Following \cite{Closset:2013sxa}, we also define
\bea\label{overline}
\overline{K} & \equiv & \frac{1}{\Omega^2}\partial_w \ = \ \frac{1}{\Omega^2}K^*~, \qquad \overline{Y} \ \equiv \ \frac{1}{sc}(\partial_{{z}} -{h}\partial_{{w}})~,
\eea
which again have natural expressions as bilinears.
The dual one-forms  to $K$ and $Y$ are
\bea
K^\flat &=& \Omega^2 (\diff w + h \diff z)~, \qquad Y^\flat \ = \ sc\,  \diff z~.
\eea
These both have Hodge type $(1,0)$ with respect to $I_+$, showing
that $z$ and $w$ are local holomorphic coordinates for this complex structure. 
In fact $K^\flat,Y^\flat$ form a basis for $\Lambda^{1,0}_+$. On the other hand 
$K^\flat,(Y^\flat)^*$ form a  basis for $\Lambda^{1,0}_-$. 
It follows that $K$ generates a complex transversely holomorphic foliation of $M_4$, 
where the transverse complex structure has opposite sign for $I_\pm$, while the 
complex structure of the leaves is the same for both $I_\pm$. In other words,
$z$ is a transverse holomorphic coordinate for $I_+$, but it is $\bar{z}$ that 
is a transverse holomorphic coordinate for $I_-$. In the quasi-regular and regular cases, 
this means that the induced complex structure on the (orbifold) Riemann surface $\Sigma_2=M_3/U(1)$ 
has the opposite sign for $I_\pm$.

Finally, let us introduce the complex two-form bilinears
\bea
\mathcal{P}_\pm & \equiv & \frac{1}{2}\zeta_\pm (\sigma_\pm)_{\mu\nu}\zeta_\pm \, \diff x^\mu\wedge \diff x^\nu~.
\eea
These are nowhere zero sections of $\Lambda^{2,0}_\pm\otimes L_\pm^2$, where $L_\pm\cong (\Lambda^{2,0}_\pm)^{-1/2}$ are spin$^c$ line bundles 
for the Killing spinors $\zeta_\pm$. We shall consider a  class of geometries in which 
the background Abelian gauge field $A_\mu$ that couples to the R-symmetry 
is real. In this case we may write
\bea
\mathcal{P}_+ &=& (\det g_{M_4})^{1/4} s\, (\diff w+h\diff z)\wedge \diff z \ = \ \Omega^3c\, \ex^{-\ii\omega}\, \diff w\wedge \diff z~,\nn\\
\mathcal{P}_- &=& (\det g_{M_4})^{1/4} \frac{\Omega^2}{s}\, (\diff w+h\diff z)\wedge \diff \bar{z} \ = \ \Omega^3c\, \ex^{\ii\omega}\, (\diff w +h\diff z)\wedge \diff \bar{z}~,
\eea
where $(\det g_{M_4})^{1/4}=\Omega^2 c$ and $s=\Omega\, \ex^{-\ii\omega}$, with $\omega$ real \cite{Assel:2014paa}. 
Notice that the latter implies $\overline{Y}=Y^*$ in (\ref{overline}), where the star denotes complex conjugation.
By definition
\bea
\diff\mathcal{P}_\pm &=& -\ii \mathcal{Q}_\pm\wedge\mathcal{P}_\pm~,
\eea
where $\mathcal{Q}_\pm$ are the associated Chern connections. We calculate
\bea\label{Qpm}
\mathcal{Q}_\pm &=& \diff^c_\pm\log (\Omega^3c) \pm \diff\omega~,
\eea
where $\diff^c_\pm \equiv \ii (\bar{\partial}^\pm-\partial^\pm)$. The background Abelian gauge field $A_\mu$
is then
\bea\label{A}
A &=& -\frac{1}{2}\mathcal{Q}_+ \ = \ \frac{1}{2}\mathcal{Q}_-~.
\eea
It follows that $\pm A$ is a connection on $\mathcal{K}_\pm^{-1/2}$, where $\mathcal{K}_\pm\equiv \Lambda^{2,0}_\pm$ is the canonical bundle for the
$I_\pm$ complex structure. Notice that $\diff A$ in fact has Hodge type $(1,1)$ for both 
$I_\pm$, and thus $\mathcal{K}_\pm$ are both holomorphic line bundles (with respect to their relevant complex structures).

\subsection{Hopf surfaces}\label{Hopfgeometry}

In most of the paper
 we will focus on backgrounds $M_4=  S^1\times M_3$, where 
 the three-manifold $M_3$ has finite fundamental group. This means that the universal
 covering space of $M_3$ is a three-sphere $S^3$, and moreover $M_3\cong S^3/\Gamma$, where 
 $\Gamma\subset SO(4)$.\footnote{This is Thurston's elliptization conjecture, now a theorem.} These
 so-called \emph{spherical three-manifolds} are classified: $\Gamma$ is either cyclic, or is a central extension of a dihedral, tetrahedral, octahedral, or icosahedral group. The cyclic case corresponds to Lens spaces $L(p,q)$, with fundamental group $\Gamma\cong \Z_p$. 
Another particularly interesting case is when $\Gamma$ is the binary icosahedral group: here $M_3$ is the famous Poincar\'e homology 
sphere. Being a homology sphere means that $\Gamma$ is a perfect group (equal to its commutator subgroup), and hence 
has trivial Abelianization. In fact $\pi_1(M_3)\cong \Gamma$ has order 120, while $H_1(M_3,\Z)$ is trivial. 
Of course our three-manifold $M_3$ also comes equipped with extra structure, and $M_4=S^1\times M_3$ must be
ambi-Hermitian with respect to $I_\pm$. As we shall see, one can realise such  supersymmetric $S^1\times M_3$  
backgrounds as \emph{Hopf surfaces}, at least for $\Gamma\subset SU(2)\subset U(2)\subset SO(4)$.

\subsubsection{Primary Hopf surfaces}\label{primaryHopfgeometry}

Let us first describe this structure in the case when $M_3\cong S^3$.  
Here $M_4$ is by definition a \emph{primary Hopf surface} -- a compact complex surface obtained as a quotient of $\C^2\setminus\{0\}$ by a free $\Z$ action. These were studied in detail in \cite{Assel:2014paa}, and in what follows we shall review and extend the analysis in this reference.

In the $I_+$ complex structure global complex coordinates
$(z_1^+,z_2^+)$ on the covering space
$\C^2\setminus\{0\}$ are expressed in terms of the local complex coordinates 
$z$, $w$ defined in the previous subsection via
\bea\label{z1z2plus}
z_1^+ &=& \ex^{|b_1|(\ii w - z)}~, \nonumber\\
z_2^+ & = & \ex^{|b_2|(\ii w + z)}~.
\eea
The Hopf surface $M_4=S^1\times S^3$ is  the quotient of $\C^2\setminus \{0\}$ by the $\Z$ action generated by
\bea\label{Zaction}
(z_1^+,z_2^+) &\rightarrow & (\mathtt{p_+}z_1^+,\mathtt{q_+}z_2^+)~,
\eea
where the complex structure parameters are $\mathtt{p_+}\equiv \ex^{\beta |b_1|}$, $\mathtt{q_+}\equiv \ex^{\beta|b_2|}$.\footnote{For a 
general primary Hopf surface these parameters may be complex.} Notice that we may equivalently 
reverse the sign of the generator in (\ref{Zaction}), with $(z_1^+,z_2^+)\rightarrow (\mathtt{p_-}z_1^+,\mathtt{q_-}z_2^+)$ 
and $\mathtt{p_-}\equiv \mathtt{p}_+^{-1}$, $\mathtt{q}_-\equiv \mathtt{q}^{-1}_+$.

We may further express these complex coordinates in 
terms of four real coordinates $\varrho,\tau,\psi_1,\psi_2$ via
\bea
w \ = \ \frac{1}{2|b_1|}\psi_1 + \frac{1}{2|b_2|}\psi_2-\ii \tau - \ii Q(\varrho)~, \quad z \ = \ u(\varrho) - \ii\left(\frac{1}{2|b_1|}\psi_1 - \frac{1}{2|b_2|}\psi_2\right)~,
\eea
where in the notation of section \ref{sec:background} we have that $Q=\ii P$ is real. 
We have introduced a polar coordinate $\varrho\in[0,1]$ on $S^3$, so that the real functions $u=u(\varrho)$, $Q=Q(\varrho)$; 
these obey equations that may be found in  \cite{Assel:2014paa}, although their precise form won't be relevant in what follows.\footnote{Compared 
to reference \cite{Assel:2014paa} we have defined $\psi_i=\mathrm{sgn}(b_i)\varphi_i$, $i=1,2$, and recall from footnote~2 
that we have also reversed the overall sign of the two complex structures $I_\pm$ compared with that reference, meaning that 
$z_i^\pm\mid_{\mathrm{here}}=\overline{z_i^\pm}\mid_{\mathrm{there}}$.}
We then have
\bea\label{zplus}
z_1^+ &=& \ex^{|b_1|\tau}\ex^{|b_1|(Q-u)}\ex^{\ii \psi_1}~, \nn\\
z_2^+ &=& \ex^{|b_2|\tau}\ex^{|b_2|(Q+u)}\ex^{\ii \psi_2}~,
\eea
and the quotient by (\ref{Zaction}) simply sets  $\tau\sim\tau+\beta$, with $\tau$ a coordinate on $S^1=S^1_\beta$. 
In \cite{Assel:2014paa} a general class of metrics on $M_3\cong S^3$ was studied, with 
 $U(1)\times U(1)$ isometry. The latter has standard generators 
$\partial_{\psi_1}$, $\partial_{\psi_2}$, and the Reeb vector field is 
\bea\label{ReebHopf}
\xi &=& \partial_\psi \ = \  |b_1|\partial_{\psi_1} + |b_2|\partial_{\psi_2}~.
\eea

The complex structure $I_-$ also equips $M_4=S^1\times S^3$ with the structure of a Hopf surface.
Global complex coordinates on the covering space $\C^2\setminus\{0\}$ are now
\bea\label{zminus}
z_1^- &=& \ex^{-|b_1|[\ii (w+2\ii Q) + \bar{z})} \ = \ \ex^{-|b_1|\tau}\ex^{|b_1|(Q-u)}\ex^{-\ii \psi_1}~, \nn\\
z_2^- &=&   \ex^{-|b_2|[\ii (w+2\ii Q) - \bar{z})} \ = \ \ex^{-|b_2|\tau}\ex^{|b_2|(Q+u)}\ex^{-\ii \psi_2}~.
\eea
In particular notice in these coordinates the complex structure parameters 
are $\mathtt{p_-}\equiv \ex^{-\beta |b_1|}=\mathtt{p_+^{-1}}$, $\mathtt{q_-}\equiv \ex^{-\beta|b_2|}=\mathtt{q_+^{-1}}$.
Notice also that $w+2\ii Q$ and $\bar{z}$ are local complex coordinates for $I_-$, 
the former following from $\diff w+2\ii \diff Q = (\diff w + h\diff z) + 2\ii\partial_{\bar{z}}Q\diff \bar{z}$, both 
of which have Hodge type $(1,0)$ with respect to $I_-$.
The  fact that $(z_1^-,z_2^-)$ cover $\C^2\setminus\{0\}$ follows from an analysis similar to that in 
\cite{Assel:2014paa} for the $I_+$ complex structure.

Another fact that we need from \cite{Assel:2014paa}, that will be particularly important 
when we come to solve globally for the modes in section \ref{sec:primary}, is that 
\bea\label{omegaS3}
\omega & =& -\psi_1 - \psi_2~.
\eea
Recall here that $s=\Omega\, \ex^{-\ii \omega}$, which for example enters the Chern connections (\ref{Qpm}), 
and hence the background R-symmetry gauge field (\ref{A}). This choice of phase in $s$ 
is fixed uniquely by requiring that $A$ is a \emph{global one-form} on $M_3\cong S^3$. The Killing spinors 
$\zeta_\pm$ are then globally defined as sections of trivial rank 2 bundles over $M_4=S^1\times S^3$. 
Gauge transformations $A\rightarrow A+\diff\lambda$ of course shift $\omega\rightarrow \omega - 2\lambda$. 

\subsubsection{Secondary Hopf surfaces}
\label{secondaryHopfgeometry}

More generally, if $M_3$ has finite fundamental group $\Gamma$ then $M_4=S^1\times M_3$ is 
a quotient of a primary Hopf surface by $\Gamma$. These are examples of \emph{secondary Hopf surfaces}.

Let us first look at cyclic $\Gamma\cong \Z_p$. 
In order that the quotient by $\Gamma$ preserves supersymmetry, in particular $s$ must be invariant. 
In terms of either $I_\pm$ complex structures, this means that $\Gamma\cong \Z_p\subset U(1)\subset SU(2)$, 
with $SU(2)$ acting on $\C^2$ in the standard two-dimensional representation $\mathbf{2}$. The generator of 
this $U(1)$ subgroup of the isometry group $U(1)\times U(1)$ is the Killing vector $\chi=\partial_{\psi_1}-\partial_{\psi_2}$, and 
from (\ref{omegaS3}) we see that $\mathcal{L}_\chi s=0$. 
It follows that $M_4=S^1\times M_3$ is isomorphic to the secondary Hopf surface $(\C^2\setminus\{0\})/(\Z\times \Z_p)$,
in both complex structures. The three-manifold $M_3$ is the Lens space $L(p,1)$ in this case. Notice 
that $\chi$ commutes with the Reeb vector field $\xi$, and hence $|b_1|,|b_2|$ 
(which determine the complex structure parameters  $\mathtt{p}_\pm$, $\mathtt{q}_\pm$) can be arbitrary.

We may also realise supersymmetric backgrounds with non-Abelian fundamental groups. 
Here we may take $\Gamma\subset SU(2)$ to act on $\C^2$ 
in the  representation $\mathbf{2}$. In order 
for $\Gamma$ to act isometrically we assume the isometry group to be enlarged to $U(2)\cong U(1)\times_{\mathbb{Z}_2} SU(2)$, with 
the Reeb vector embedded along $U(1)$. This means $|b_1|=|b_2|$. The metric on $M_3\cong S^3$ is then 
that of a Berger sphere
\bea
\diff s^2_{M_3} &=& \diff\theta^2+\sin^2\theta\diff\varphi^2 + v^2(\diff\varsigma+\cos\theta\diff\varphi)^2~,
\eea
where $v>0$ is a squashing parameter, and $\varsigma=\psi_1+\psi_2$, $\varphi=\psi_1-\psi_2$. This special case of a Hopf surface background was studied 
in appendix C of \cite{Assel:2014paa}, and has $b_1=-b_2=1/2v$, and $I_+$ complex coordinates
\bea
z_1^+ &=& \sqrt{2}\, \ex^{\frac{\tau}{2v}}\cos\frac{\theta}{2}\ex^{\ii \psi_1}~, \qquad z_2^+ \ = \  \sqrt{2}\, \ex^{\frac{\tau}{2v}}\sin\frac{\theta}{2}\ex^{\ii \psi_2}~.
\eea
In particular $|z_1^+|^2+|z_2^+|^2 = 2\ex^{\tau/v}$ is invariant under $SU(2)$. The $I_-$ complex coordinates are
\bea
(z_1^-,z_2^-) &=& \ex^{-\frac{\tau}{v}}\, (z_1^+,z_2^+)^*~,
\eea
meaning that the $SU(2)$ group acts in the complex conjugate representation $\bar{\mathbf{2}}$ in the $I_-$ complex 
structure. As is well known, $\mathbf{2}\cong \bar{\mathbf{2}}$, and thus again $M_4=S^1\times M_3$ is isomorphic to the secondary Hopf surface $(\C^2\setminus\{0\})/(\Z\times \Gamma)$ in both complex structures. 
Of course finite subgroups 
$\Gamma\subset SU(2)$ have an ADE classification, where the A series are precisely the 
Abelian $\Gamma\cong\Z_p$ quotients of primary Hopf surfaces described at the beginning of this subsection, 
while the D and E groups are the dihedral series and tetrehedral $E_6$, octahedral $E_7$ and icosahedral $E_8$ groups, respectively. 

We may also describe the complex geometry of the associated Hopf surfaces algebraically. 
Consider the polynomials
\bea\label{hsequations}
f_{A_{p-1}} &=& Z_1^p+Z_2^2+Z_3^2~, \qquad f_{D_{p+1}} \ = \ Z_1^p + Z_1Z_2^2 + Z_3^2~,\\
f_{E_6} &=& Z_1^3+Z_2^4+Z_3^2~, \qquad f_{E_7} \ = \  Z_1^3+Z_1Z_2^3+Z_3^2~,\qquad f_{E_8} \ =\  Z_1^3+Z_2^5+Z_3^2~,\nn
\eea
on $\C^3$ with coordinates $(Z_1,Z_2,Z_3)$. The zero sets
\bea
X &\equiv & \{f(Z_1,Z_2,Z_3) \ = \ 0\}\  \subset \  \C^3~,
\eea
have an isolated singularity at the origin $o$ of $\C^3$. These are all weighted homogeneous 
hypersurface singularities, meaning they inherit a $\C^*$ action from the $\C^*$ action 
$(Z_1,Z_2,Z_3)\rightarrow (\Nlambda^{w_1}Z_1,\Nlambda^{w_2}Z_2,\Nlambda^{w_3}Z_3)$ on
$\C^3$, where $w_i\in\mathbb{N}$ are the weights, $i=1,2,3$, and $\Nlambda\in\C^*$. 
For example, $f_{A_{p-1}}$ has degree $d=2p$ under the weights $(w_1,w_2,w_3)=(2,p,p)$, 
while $f_{E_8}$ has degree $d=30$ under the weights $(w_1,w_2,w_3)=(10,6,15)$. 
The smooth locus $X\setminus\{o\}\cong \R\times M_3$, where $M_3=S^3/\Gamma_{ADE}$, 
while the quotients $(X\setminus\{o\})/\Z$ are precisely the ADE secondary Hopf surfaces
described above. Here $\Z\subset\C^*$ is embedded as $n\rightarrow \Nlambda^n$
for some fixed $\Nlambda>1$. The Reeb vector field action is quasi-regular, generated by 
$\Nlambda\in U(1)\subset \C^*$. The quotient $\Sigma_2=M_3/U(1)$ is in general 
an orbifold Riemann surface of genus $g=0$.

\subsection{Flat connections}\label{sec:flat}

The path integral of any four-dimensional $\mathcal{N}=1$ theory with an R-symmetry 
on one of the supersymmetric backgrounds $S^1\times M_3$ of section \ref{sec:background} localizes. In particular, 
the supercharges generated by $\zeta_\pm$ localize the vector multiplet 
onto instantons and anti-instantons, respectively \cite{Assel:2014paa}, which intersect
on the \emph{flat connections}. In the Hamiltonian formalism for computing the 
supersymmetric Casimir energy, we will  then need to study flat connections on the 
covering space $\R\times M_3$. The two spaces $S^1\times M_3$ and $\R\times M_3$
have respectively $\tau$ periodic with period $\beta$, and $\tau\in\R$.

Recall that flat connections on $M_4$ with gauge group $G$ are in one-to-one correspondence with 
\bea\label{flatconnections}
\mathrm{Hom}(\pi_1(M_4)\rightarrow G)/\mathrm{conjugation}~.
\eea
In particular a flat $G$-connection is determined by its holonomies, which define a homomorphism 
$\varrho:\pi_1(M_4)\rightarrow G$, while gauge transformations act by conjugation. 
In the path integral on $M_4=S^1\times M_3$ we have $\pi_1(M_4)\cong \Z\times \pi_1(M_3)$, 
with $\pi_1(S^1)\cong \Z$. A flat connection is then the sum of pull-backs 
of flat connections on $S^1$ and $M_3$, and we denote the former by 
 $\mathcal{A}_0$.
On the other hand in the Hamiltonian formalism instead 
$M_4=\R\times M_3$, so that $\pi_1(M_4)\cong \pi_1(M_3)$, and a
flat connection on $M_4$ is simply the pull-back of a flat connection on $M_3$.

When $\pi_1(M_3)$ is finite, which is the case for the primary and secondary Hopf surfaces in 
section \ref{Hopfgeometry}, the number of inequivalent flat connections on $M_3$
is also finite. The path integral on $S^1\times M_3$ correspondingly
splits into a finite sum over these topological sectors, together with a matrix integral over 
the holonomy of $\mathcal{A}_0$.
In the Hamiltonian formalism on $\R \times M_3$, instead for each 
flat connection on $M_3$ we will obtain a different supersymmetric quantum mechanics on $\R$.

A matter multiplet will be in some representation $\mathcal{R}$ of the gauge group $G$. 
In the presence of a non-trivial flat connection on $M_4=\R\times M_3$, this 
matter multiplet will be a section of the associated flat vector bundle, tensored 
with $\mathcal{K}_+^{-\qq/2}$ if the matter field has R-charge $\qq$. 
The latter follows since recall that the background R-symmetry 
gauge field $A$ is a connection on $\mathcal{K}_+^{-1/2}\cong \mathcal{K}_-^{+1/2}$.
For the Hopf surface cases of interest this will always be a trivial bundle, albeit with a generically non-flat connection, and we 
hence suppress this in the following discussion. 
Concretely then, composing $\varrho:\pi_1(M_4)\rightarrow G$ with the representation 
$\mathcal{R}$ of $G$ determines a corresponding flat connection in the 
 representation $\mathcal{R}$, and the scalar field in the matter multiplet is a section 
of the vector bundle
\bea\label{flatmatter}
\mathcal{V}_{\mathrm{matter}} &=& (\widetilde{M_4}\times V)/\pi_1~.
\eea
Here $\widetilde{M_4}$ is the universal cover of $M_4$ (which is $\R\times S^3$ for Hopf surfaces), 
$\pi_1=\pi_1(M_4)$ is the fundamental group, $V\cong \C^M$ is the vector space associated to $\mathcal{R}$, 
and the action of $\pi_1$ on $V$ is determined by the flat $\mathcal{R}$-connection described above. 
The scalar field in the matter multiplet is then a section of the bundle (\ref{flatmatter}), 
which is a $\C^M$ vector bundle over $M_4$.

To illustrate, let us focus on the simplest non-trivial 
example, namely the Lens space $M_3=S^3/\Z_p=L(p,1)$. 
For a $G=U(1)$ gauge theory the flat connections on $\R\times M_3$ may be labelled by 
an integer $0\leq m<p$, which determines the holonomy
\bea\label{flatAbelian}
\exp\left(\ii\int_\gamma \mathcal{A}\right) &=& \ex^{2\pi \ii m/p}~.
\eea 
Here $\mathcal{A}$ is the dynamical $U(1)$ gauge field, while the circle $\gamma$ generates $\pi_1(\R\times M_3)\cong \Z_p$. 
The associated homomorphism $\varrho:\Z_p\rightarrow U(1)$ is generated by  
$\varrho(\omega_p)=\omega_p^m$, where $\omega_p\equiv \ex^{2\pi \ii/p}$ is 
a  primitive $p$th root of unity.
For a $U(N)$ gauge theory the flat connections are similarly labelled by $0\leq m_i<p$, where $i=1,\ldots,N$ 
runs over the generators of the Cartan $U(1)^N$ subgroup of $U(N)$.
 These are permuted by the Weyl group, 
so without loss of generality one may choose to order $m_1\leq m_2\leq \cdots\leq m_N$, 
and label the flat $U(N)$ connection by  a vector $\mathbf{m}=(m_1,\ldots,m_N)$. 
Now $\varrho:\Z_p\rightarrow U(N)$ is generated by 
$\varrho(\omega_p)=\mathrm{diag}(\omega_p^{m_1},\ldots,\omega_p^{m_N})\in U(N)$.

An irreducible representation of $U(1)$ is labelled by the charge $\anothers\in\Z$, so $\mathcal{R}=\mathcal{R}_\anothers$.  
In the presence of the flat connection (\ref{flatAbelian}), a matter field 
in this representation becomes a section of the 
 line bundle $L$ over $\R\times L(p,1)$ 
with first Chern class $c_1(L)\in H^2(\R\times L(p,1),\Z)\cong \Z_p$ given by
$c_1(L)\equiv \anothers m$ mod $p$. 
Equivalently, on the universal covering space $\widetilde{M}_4\cong 
\R\times S^3\cong\C^2\setminus\{0\}$, the relevant sections of $\mathcal{V}_{\mathrm{matter}}$
may be identified with functions on $\widetilde{M}_3$
which pick up  a phase $\ex^{2\pi \ii c_1(L)/p}$ under the generator 
of the $\Z_p$ action. More generally, for a $U(N)$ gauge group 
we may decompose the representation $\mathcal{R}=\oplus_\rho V_\rho$
into weight spaces, with weights $\rho$. This then essentially reduces 
to the line bundle case above, with the part of the matter field in
$V_\rho$ now being a section of $L$ with $c_1(L)\equiv \rho(\mathbf{m})$ mod $p$. 
For example, the fundamental representation of $U(N)$ has weights 
$\rho_i(\mathbf{m})=m_i$, $i=1,\ldots,N$, the adjoint representation
has weights $\rho_{ij}=m_i-m_j$, {\it etc}.

\section{Supersymmetric Casimir energy}
\label{sec:Casimir}

In this section we review the two  approaches to define the supersymmetric Casimir energy $\Esusy$, involving the path integral formulation 
on a compact manifold $S^1\times M_3$, and the Hamiltonian formalism on its covering space $\R\times M_3$, respectively. We also 
 present a geometric interpretation of the shortening conditions 
previously discussed in~\cite{Closset:2013sxa,Assel:2015nca}.

\subsection{Path integral formulation}

On general grounds \cite{Closset:2013vra}, the localized path integral of a four-dimensional $\mathcal{N}=1$ theory with an R-symmetry on 
$M_4=S^1\times M_3$ is expected to depend on the background geometry only via the complex structure(s) of $M_4$. 
For example, for the primary Hopf surfaces described in section \ref{primaryHopfgeometry} the  complex
structure parameters are  $\mathtt{p}_\pm=\ex^{\pm\beta|b_1|}, \mathtt{q}_\pm=\ex^{\pm\beta|b_2|}$, which may equivalently be thought of as 
specified by the choice of Reeb vector field $\xi$ in (\ref{ReebHopf}) (together with $\beta$). For a secondary Hopf surface $S^1\times M_3$, the localized partition function also carries information about the finite fundamental group $\Gamma=\pi_1(M_3)$. 
Of course the partition function will also depend on the choice of $\mathcal{N}=1$ theory, 
through the choice of gauge group, matter representation, and in particular on the R-charges of the matter fields.

In analogy with the usual zero point energy of a field theory,  the supersymmetric Casimir energy was defined in \cite{Assel:2014paa} 
as a limit of  the supersymmetric partition function  $Z^\mathrm{susy}_{S^1_\beta \times M_3}$, namely the path integral with periodic boundary conditions
for the fermions along  $S^1_\beta$. More precisely, 
\be
 E_{\rm susy}  \ \equiv  \  - \lim_{\beta \to \infty} \frac{\diff }{\diff \beta} \log Z^\mathrm{susy}_{S^1_\beta \times M_3}~.
 \label{casimpi}
\ee
This may be computed using localization. As already mentioned in section \ref{sec:flat}, the 
vector multiplet localizes onto flat connections  for the gauge group $G$, while 
at least for primary Hopf surfaces the matter multiplet localizes to zero. 
The localized partition function  comprises the contributions of one-loop determinants for the vector and chiral multiplets of the theory, evaluated around each such BPS locus, and one then integrates/sums over the space of flat connections. 
For primary Hopf surfaces  ($M_3\cong S^3$), the only non-trivial gauge field holonomy is for the flat connection $\mathcal{A}_0$ along $S^1_\beta$ \cite{Assel:2014paa}. On the other hand, if $\pi_1(M_3)$ is non-trivial
one should also sum or integrate over flat connections on $M_3$, in the cases that $\pi_1(M_3)$ is finite, or infinite, respectively \cite{Nishioka:2014zpa}.  

For  primary Hopf surfaces the partition function factorises $Z^\mathrm{susy}_{S^1_\beta \times S^3} = \me^{-\beta E_\mathrm{susy} (|b_1|,|b_2|)} {\cal I}$, where ${\cal I}$ is a matrix integral over the gauge field holonomies on $S^1_\beta$,  known as the supersymmetric index \cite{Romelsberger:2005eg}.  The latter  does not contribute to the limit (\ref{casimpi}),  and thus in order to compute  $E_{\rm susy}$  one can effectively set the gauge field ${\cal A}_0=0$
in the one-loop determinants.  The regularization of these determinants is rather delicate and it was proved
in \cite{Assel:2015nca} that  regularizations respecting supersymmetry give rise to a partition function with large and small 
$\beta$ limits consistent with general principles \cite{DiPietro:2014bca}.  See appendix C of   \cite{Assel:2015nca}. 

For secondary Hopf surfaces the partition function is a sum of contributions over sectors with a fixed flat connection on $M_3$. 
Let us label these sectors as $\iflat\in\Mflat$. Recall that in the special case that $M_3=L(p,1)=S^3/\Z_p$ is a Lens space and 
$G=U(N)$ we may identify $\Mflat$ with the space of vectors $\mathbf{m}=(m_1,\ldots,m_N)$, where 
$0\leq m_i<p$ and $m_1\leq m_2\leq \cdots\leq m_N$.
 Then from the definition (\ref{casimpi}) it is clear that the supersymmetric Casimir energy is given by 
\bea
E_{\mathrm{susy}} &=& \min_{\iflat\in\Mflat}\ \{E_{\mathrm{susy},\iflat}\}~,
\eea
where for each $\iflat$  we have defined  a  ``supersymmetric Casimir energy in the sector $\iflat$'' as 
\bea
E_{\mathrm{susy},\iflat} &=& -\lim_{\beta\rightarrow \infty} \frac{\diff}{\diff\beta} \log Z_{\iflat}~.
\eea
In the Lens space case $M_3=L(p,1)= S^3/\Z_p$ the partition functions $ Z_{\iflat}$, which include the Casimir contributions
 $ E_{\mathrm{susy},\iflat}$, have been computed in \cite{Nishioka:2014zpa}.  

\subsection{Hamiltonian formulation}

Because the geometries of interest are of the form $M_4=S^1\times M_3$, with $\de_\tau$ the Killing vector 
generating translations on $S^1$, we can  consider the theories on the covering space $M_4= \R\times M_3$, 
 employing the Hamiltonian formalism.\footnote{On  $M_4= \R\times M_3$ one usually works in Lorentzian signature. In this paper, however, we will always remain in Euclidean signature. One can then take the point of view that the Wick rotation ($t=\ii\tau$) to pass from Euclidean to Lorenztian signature 
can be done after the reduction to one dimension. 
In practice, we will never need to perform this last step.}
 These two approaches have been shown to yield equivalent results for both the supersymmetric Casimir energy, as well as the index ${\cal I}$,
for primary Hopf surfaces, $M_3\cong S^3$.
It was argued in  \cite{Romelsberger:2005eg} that the  supersymmetric index cannot depend on continuous couplings of the theory or the RG scale, and therefore may be  computed in the free limit (assuming this exists). 
We return to discussing the supersymmetric index  in Appendix \ref{letters:sec}. The supersymmetric Casimir energy can also be obtained as the vacuum expectation value of the 
supersymmetric  (Weyl ordered)  Hamiltonian $H_\mathrm{susy}$, and again it can be reliably computed in a free theory
   \cite{Assel:2015nca}. This can be further Kaluza-Klein reduced on $M_3$ to give a supersymmetric quantum mechanics on $\R$, with an infinite number of fields, 
   organised into multiplets of one-dimensional supersymmetry. 
Then $E_{\mathrm{susy}}=\langle H_{\mathrm{susy}}\rangle$,  where $H_{\mathrm{susy}}$ is the total Hamiltonian for this  supersymmetric quantum mechanics. 
If supersymmetric regularizations are employed, then this definition has been shown to agree with  (\ref{casimpi}) in the primary Hopf 
surface case $M_3\cong S^3$ 
\cite{Assel:2015nca}.  

This formalism can also be utilised when  $\pi_1(M_3)$  is non-trivial (and  finite), as we will see in more detail  later in the paper. 
 In this case there is a  supersymmetric quantum mechanics for 
\emph{each} flat connection on $M_3$.  This leads to a definition of  ``supersymmetric Casimir energy in the sector $\iflat$'' that will depend on 
the flat connection $\iflat\in\Mflat$, thus 
  \bea
  E_{\mathrm{susy},\iflat} &=& \langle H_{\mathrm{susy},\iflat}   \rangle ~.
  \eea 
 We will see that this quantum-mechanical definition of    $E_{\mathrm{susy},\iflat}$ coincides with the path integral definition given previously, 
in any sector $\iflat=\mathbf{m}$, for Lens space secondary Hopf surfaces with $M_3=L(p,1)=S^3/\Z_p$.  Of course, the actual supersymmetric Casimir energy of the theory 
will be given by the minimum  $E_{\mathrm{susy},\iflat}$ among all flat connections.

In the simplest case, where $M_3=S^3_\mathrm{round}$, the Hamiltonian formalism can be used to obtain explicitly all of the 
 modes and their eigenvalues \cite{Gerchkovitz:2013zra,Lorenzen:2014pna,Assel:2015nca}.
Only a subset of \emph{unpaired} modes contribute to $E_{\mathrm{susy}}$ 
\cite{Lorenzen:2014pna}. These modes where shown in 
\cite{Assel:2015nca} to correspond to \emph{short} 1d supersymmetry multiplets (chiral and Fermi multiplets). This feature  extends to more general geometries, 
where the unpaired modes obey shortening conditions   taking the form of linear first order differential equations \cite{Closset:2013sxa}.  

\subsection{Twisted variables}

In what follows we will focus attention on a chiral multiplet.
Using a    set of ``twisted variables'' \cite{Closset:2013sxa},
the fermion of a chiral multiplet can be replaced by  a pair of anticommuting complex scalar
 fields $B$ and $C$. 
 Thus such a multiplet comprises the four scalar fields $(\phi,B,C,F)$, with R-charges  $(r,r-2,r,r-2)$, respectively. 
  There  is also a set of tilded fields $(\tilde \phi,\tilde B,\tilde C,\tilde F)$ with opposite sign R-charges, that 
   are eventually simply  related to the untilded  fields by complex conjugation. 
  The  localizing deformation\footnote{This coincides with the standard chiral multiplet Lagrangian for 
  a particular choice of the parameter $\kappa$.}  in these variables takes the  simple form 
\bea
{\cal L}_\mathrm{loc} & = & 4 \tilde \phi \Db  \phi + 2 \tilde \Psi \Df \Psi  - \tilde F F~,
\label{Lloc}
\eea
where  $\tilde \Psi = (\tilde B, \tilde C)$,  $\Psi = (B,C)^T$, and  we have defined the operators
\bea
\Db &\equiv & -(\hL_{\bK} \hL_K + \hL_Y \hL_{\bY})~,\qquad \Df \ \equiv\ \ii \left(\begin{array}{cc} \hL_K &  \hL_{\bY} \\ 
- \hL_Y &  \hL_{\bK}\end{array}\right)~,
\eea
with  the first order operators
\bea
\hat {\cal L}_U & = & U^\mu (\de_\mu - \ii \qq A_\mu -\ii \mathcal{A}_\mu)~.
\label{defLs}
\eea
Here $U$ is one of the four complex vector fields $K,\bK, Y, \bY$, defined in section \ref{sec:background}, 
 $\qq$ is the R-charge of the field on which the operator is acting, and $\mathcal{A}_\mu$ denotes 
 the localized flat gauge connection, acting on the field in the appropriate representation $\mathcal{R}$. 
  As discussed in section \ref{sec:flat}, such matter fields may equivalently be identified 
  with functions on the covering space that transform appropriately under the 
  action of $\pi_1=\pi_1(M_3)$ determined by the flat connection $\mathcal{A}_\mu$. 
  This action commutes with $K$ and $\bar{K}$, as was necessary to preserve supersymmetry.
We note the following relations  
\bea\label{commutators}
[\hL_K,\hL_{\bK}] &=& 0~, \qquad [\hL_K,\hL_Y] \ = \ 0~, \qquad [\hL_K,\hL_{\bY}] \ = \ 0~.
\eea
These were proven in  \cite{Closset:2013sxa} in a fixed (local) R-symmetry gauge where $s=s(z,\bar z)$ (and 
without the flat connection), although 
it is obvious that they are valid in any gauge. In particular they are valid in the 
unique global non-singular gauge (\ref{omegaS3}), relevant for Hopf surfaces. 

 The unpaired modes were shown in \cite{Closset:2013sxa} to satisfy the shortening  conditions 
\bea
\hat{\mathcal L}_Y \Bold \ = \ 0 \ , && \qquad \qquad \ii \hat{\mathcal L}_K \Bold \ = \ \lambda^B \Bold \ ,
\label{1stOrderConditionsB}\nonumber\\
\hat{\mathcal L}_{\overline Y}\, \phiold \ = \ 0 \ , && \qquad \qquad  \ii \hat{\mathcal L}_K \phiold \ = \ \lambda^\phi \phiold \, ,
\label{1stOrderConditionsphi}
\eea
where we have denoted the modes  $\Bold$, $\phiold$, to distinguish them from the closely related modes to be introduced momentarily.
 It is worth emphasizing that these equations are valid both on $S^1\times M_3$ and on $\R \times M_3$; however, the eigenvalues $\lambda^B$, $\lambda^\phi$ are different in the 
two cases. In particular, on $S^1\times M_3$ one expands all fields in Kaluza-Klein modes over $S^1$ \cite{Assel:2014paa}, thus $\Phiold (\tau,x_i)= \Phiold (x_i) \, \me^{-\ii n\tau }$, where $n\in \Z$ and  $x_i$, $i=1,2,3$ are coordinates on $M_3$. Correspondingly we have
$\lambda^\Phi = -\frac{\ii }{2} n + \lambda^\Phi_\xi $, where $\lambda^\Phi_\xi \in \R$ is the Reeb charge of the modes on $M_3$:  
\be\ii \hL_{\frac{1}{2}\xi} \Phiold (x_i) \ = \ \lambda^\Phi_\xi \Phiold (x_i)~.\ee
  On the other hand, using  the equations  (\ref{1stOrderConditionsphi}) in the context of the Hamiltonian formalism on $\R \times M_3$   \cite{Assel:2015nca}, one has effectively to set $n=0$, and therefore   in this case  $\lambda^\Phi =  \lambda^\Phi_\xi$. 

In order to compute $E_\mathrm{susy}$ in principle one should consider the Hamiltonian canonically conjugate to (\ref{Lloc}), insert all modes obeying their (free) equations of motion, and then reduce the problem to one dimension  \cite{Lorenzen:2014pna}.  Alternatively, one can focus on the unpaired modes, giving rise to short 1d multiplets, and determine their $\Sigma$-charge, for example by  analysing  the reduced supersymmetry transformations \cite{Assel:2015nca}.  Here  
$\Sigma$ is the Hermitian operator appearing in the one-dimensional supersymmetry algebra 
\bea\label{model}
&&\{Q,Q^\dagger \} \ = \ 2(H_\mathrm{susy}-\Sigma)~,\qquad Q^2 \ =\ 0\ , \nn \\ [2mm]
&&[H_\mathrm{susy}, Q]\ =\ [\Sigma, Q]\ =\ 0~.
\eea
Then $E_\mathrm{susy}$ is determined using the fact that for every multiplet $\langle H_\mathrm{susy}\rangle = \langle \Sigma \rangle $ \cite{Assel:2015nca}.

\subsection{Unpaired modes on $\R\times M_3$}\label{sec:unpaired}

In the path integral formalism, localization reduces the problem to computing the one-loop determinant associated to (\ref{Lloc}). Correspondingly,  in the Hamiltonian formulation, we   consider modes obeying the equations of motion following from (\ref{Lloc}), namely 
\bea
\Db \phi &=& 0~, \qquad \qquad \Df \Psi \ = \ 0~.
\label{physops}
\eea
It is simple to show that  modes satisfying the equations in (\ref{physops}) are paired by supersymmetry.
Indeed, if $\phi$ is a bosonic zero mode, $\Db\phi=0$,  one can check using (\ref{commutators}) that $\Psi = ( \hL_{\bY}\phi,  -\hL_K\phi)^T$ 
is a fermionic zero mode, so $\Df \Psi=0$. Conversely, if $(B,C)^T$ is a fermionic zero mode, $\Df (B,C)^T=0$, 
one can check that $\phi\equiv C$ is a bosonic zero mode, so $\Db\phi=0$. 
Modes that are paired this way form long multiplets that do not contribute 
to the supersymmetric Casimir energy.  Notice  that a fermionic zero mode satisfies
\bea\label{fermi}
\hL_{\bY} C &=& -\hL_K B~, \qquad \hL_{\bK} C \ = \ \hL_Y B~.
\eea
The  net contribution to $\Esusy$ comes from unpaired modes. These are bosonic/fermionic zero modes for which the putative 
fermionic/bosonic partner is identically zero. Thus these are 
fermionic $(B,0)$ modes satisfying (using  (\ref{fermi}))
\bea\label{Bzeromode}
\hL_Y B &=& 0 \ = \ \hL_K B~,
\eea
and 
bosonic $\phi$ modes satisfying
\bea\label{phizeromode}
\hL_{\bY}\phi &=& 0 \ = \  \hL_K\phi~.
\eea 

Recalling the definition (\ref{defLs}) and using the preliminaries in section \ref{sec:background}, one recognises the two
operators $\hL_Y $ and $\hL_K $ as the components of the twisted $\bar{\partial}^+_{A,\, \mathcal{A}} $ differential. This   denotes the $(0,1)_+$ part 
of $\diff - \ii \qq A -\ii\mathcal{A} $, where the twisting is determined by the R-symmetry  connection $A$ in (\ref{A}) and flat 
connection $\mathcal{A}$.
 In particular, the unpaired $B$ modes in (\ref{Bzeromode}) obey
\bea
\bar{\partial}^+_{A,\, \mathcal{A}} B & = & 0~,
\eea
and are therefore (twisted) holomorphic in the $I_+$ complex structure. 
Similarly, one can show the unpaired $\phi$ modes in  (\ref{phizeromode}) satisfy
\bea
\bar{\partial}^-_{A,\, \mathcal{A}} \phi & = & 0~,
\eea
where $\bar{\partial}^-_{A,\, \mathcal{A}} $ denotes the $(0,1)_-$ part 
of $\diff - \ii \qq A -\ii\mathcal{A}  $, and are therefore  (twisted) holomorphic in the $I_-$ complex structure.  Notice that more precisely the unpaired $B$ and $\phi$ 
modes are sections of  $\mathcal{V}_{\mathrm{matter}}\otimes\mathcal{K}_+^{-(r-2)/2}$ and   $\mathcal{V}_{\mathrm{matter}}\otimes\mathcal{K}_-^{r/2}$, respectively, 
where $\mathcal{V}_{\mathrm{matter}}$ is the flat matter vector bundle (\ref{flatmatter}). 

It is simple to see that the above holomorphic modes may be decomposed into modes on $M_3$ 
which 
 have definite charge under the
(twisted) Reeb vector field  $\ii \hL_{\frac{1}{2}\xi} $.
In particular,  writing a mode as 
\bea
\Phi (\tau, x_i) &=& \ex^{-2\lambda^\Phi_\xi \tau} \Phiold (x_i)~,
\label{conemodes}
\eea
and using $\ii \hL_K =  {\cal L}_{\frac{1}{2}\de_\tau} +  \ii   \hL_{\frac{1}{2}\xi} $,
one sees that 
\bea
\hL_K \Phi (\tau,x_i) & = & 0 \qquad \Longleftrightarrow \qquad  \ii \hL_{\frac{1}{2}\xi}  \Phiold (x_i) \ = \  \lambda^\Phi_\xi \, \Phiold (x_i)~.
\eea
This  shows that the modes  on $\R \times M_3$ defined by (\ref{1stOrderConditionsphi}) were indeed
 independent of  $\tau$, and therefore defined on $M_3$, as already remarked  below equation (\ref{1stOrderConditionsphi}).
Thus we can  think of the modes  (\ref{conemodes}) as the ``lifting to the cone'' of the modes
in the previous section. In fact setting  $r=\me^{-\tau}$ one sees that the metric on $\R\times M_3$ is conformally related to the metric on the cone $C(M_3)$: 
$g_{C(M_3)} = \diff r^2 + r^2 g_{M_3}$. 
Notice also that upon the Wick rotation $t=\ii\tau$, these become
 $\Phi (t, x_i) = \ex^{2\ii \lambda^\Phi_\xi t} \Phiold (x_i)$, as expected for modes solving the free equations of motion on $\R\times M_3$ in 
 Lorentzian signature  \cite{Lorenzen:2014pna}. These have to be contrasted with the modes on $S^1\times M_3$ 
 discussed earlier, namely    $\Phiold (\tau,x_i)=  \me^{-\ii n\tau } \Phiold (x_i) $.

Recall that the supersymmetry algebra acting on fields contains the anti commutation relation \cite{Dumitrescu:2012ha}
\bea
\{ \delta_{\zeta_+} ,  \delta_{\zeta_-} \} & = & 2\ii \hL_K \ =  \ 2\left( {\cal L}_{\frac{1}{2}\de_\tau} +  \ii   \hL_{\frac{1}{2}\xi} \right) ~,
\eea
where $\delta_{\zeta_\pm}$ denote supersymmetry variations with respect to the  $\zeta_\pm$ Killing spinors, respectively. 
Comparing this with  the anti-commutator in (\ref{model}),  one can identify the eigenvalues of the quantum mechanical 
operators $H_\mathrm{susy}$ and $\Sigma$ with  those of the operators\footnote{After performing the Wick rotation $t=\ii\tau$ to go to Lorentzian signature.} 
$ {\cal L}_{\frac{1}{2}\de_\tau}$  and   $- \ii \hL_{\frac{1}{2}\xi}$, acting on the classical modes, respectively. 
Therefore,  the condition  $\hL_K \Phi =0$  obeyed by the holomorphic  modes (on the cone)
 may be interpreted as showing that  the Hamiltonian eigenvalues are equal to their Reeb charge, 
 and is the counterpart of  $\langle H_\mathrm{susy}\rangle = \langle \Sigma \rangle $ in the supersymmetric quantum mechanics.

To summarise, the supersymmetric Casimir energy is computed by 
summing the Reeb charges of  (twisted) holomorphic modes on 
$\R \times M_3$,  with fermionic and bosonic modes corresponding to each  complex structure $I_\pm$, respectively.

\section{Primary Hopf surfaces}
\label{sec:primary}

In this section we re-examine the supersymmetric Casimir energy for the primary Hopf surfaces
$S^1\times S^3$ in the above formalism. This 
was first defined and computed in the path integral approach in \cite{Assel:2014paa}. 
Since $M_3\cong S^3$ there are no flat connections on $M_3$.

\subsection{Solving for the unpaired modes}\label{sec:unpairedprimary}

Recall that the unpaired $B$ and $\phi$ modes, that contribute to the supersymmetric Casimir energy, 
are zero modes on $\C^2\setminus\{0\}$ of the twisted holomorphic differentials $\bar{\partial}^+_A$, $\bar{\partial}^-_A$, 
respectively, where the background R-symmetry gauge field $A$ is given by (\ref{A}) and 
the operators are understood to act on fields of R-charge $\qq$. 
The curvature of $A$ has Hodge type $(1,1)$ with respect to both $I_\pm$, and thus 
both differentials are nilpotent.

Using the global complex coordinates defined in section \ref{primaryHopfgeometry}, it is straightforward 
to solve explicitly for these zero modes. In what follows we assume that we are working in a weight space 
decomposition of the matter representation $\mathcal{R}$, so that for a fixed weight $\rho$ we have
$B=B_\rho$ is a single scalar field.
For the unpaired $B$ modes we first note from (\ref{A}) that the 
$(0,1)_+$ part of $A$ is
\bea
A_{(0,1)_+} &=& -\frac{\ii}{2}\bar\partial^+ \log (\Omega^3 c) - \frac{1}{2}\bar\partial^+\omega~.
\eea
The equation $\bar{\partial}^+_A B=0$ may thus be rewritten as
\bea\label{BzeroHopf}
\bar{\partial}^+\left[(\Omega^3 c)^{-\qq/2}|z_1^+z_2^+|^{\qq/2} B\right] &=& 0~.
\eea
In particular notice that we have used
\bea
(z_1^+z_2^+)^{\qq/2} &=& |z_1^+z_2^+|^{\qq/2}\, \ex^{-\ii (\qq/2)\omega}~,
\eea
where $\omega=-\psi_1-\psi_2$. Recall that $\Omega$ is globally a nowhere zero function, while 
near the complex axes ({\it i.e.} $z_1^+=0$ and $z_2^+=0$) the real function $c$ behaves to leading order as 
$|c|\sim |z_1^+|$, $|c|\sim |z_2^+|$, respectively. This is required for regularity of the metric~\cite{Assel:2014paa}. 
It follows that the factor in front of $B$ inside the square bracket in (\ref{BzeroHopf}) is a real nowhere zero 
function on $\C^2\setminus\{0\}$. A basis of regular solutions is hence
\bea\label{Bm}
B &=& B_{n_1,n_2} \ \equiv \ \left(\frac{\Omega^3c}{|z_1^+z_2^+|}\right)^{\qq/2}({z}^+_1)^{n_1}({z}_2^+)^{n_2}~,
\eea
where $n_1,n_2\in \Z_{\geq 0}$. 

One may similarly solve for the unpaired $\phi$ zero modes. Since from (\ref{A}) we now have
\bea
A_{(0,1)_-} &=& \frac{\ii}{2}\bar\partial^- \log (\Omega^3 c) - \frac{1}{2}\bar\partial^-\omega~,
\eea
and one obtains a basis of regular solutions given by
\bea\label{phim}
\phi &=& \phi_{n_1,n_2} \ \equiv \ \left(\frac{\Omega^3c}{|z_1^-z_2^-|}\right)^{-\qq/2}({z}^-_1)^{n_1}({z}_2^-)^{n_2}~.
\eea

The prefactors in front of the holomorphic monomials in the modes (\ref{Bm}), (\ref{phim}) also have a simple geometric 
interpretation. Recall that the Hermitian structure $(g_{M_4},I_+)$ equips $\C^2\setminus\{0\}$ with 
the $(2,0)_+$-form
\bea
\mathcal{P}_+ &\equiv & \frac{1}{2}\zeta_+ (\sigma_+)_{\mu\nu}\zeta_+\, \diff x^\mu\wedge \diff x^\nu \ = \ 
\Omega^3 c \, \ex^{-\ii\omega}\diff w\wedge \diff z~.
\eea
On the other hand, $\C^2$ has the global holomorphic $(2,0)_+$-form\footnote{This is not to be confused 
with the conformal factor $\Omega$, especially in the following formulae.}
\bea
\Omega_+ &\equiv &  \frac{1}{2|b_1||b_2|}\diff z_1^+\wedge \diff z_2^+  \ = \ \ii z_1^+z_2^+\, \diff w\wedge \diff z~,
\eea
where we have used (\ref{z1z2plus}). Then
\bea
\frac{\Omega^3 c}{|z_1^+z_2^+|} & = & \left|\frac{\mathcal{P}_+}{\Omega_+}\right|~,
\eea
is simply the modulus of the ratio of these two canonically defined $(2,0)_+$-forms. 
A similar computation shows that
\bea
\frac{\Omega^3 c}{|z_1^-z_2^-|} & = & \left|\frac{\mathcal{P}_-}{\Omega_-}\right|~,
\eea
where we define
\bea
\Omega_- & \equiv &  \frac{1}{2|b_1||b_2|}\diff z_1^-\wedge \diff z_2^-  \ = \ \ii z_1^-z_2^-\, (\diff w+h\diff z)\wedge \diff \bar{z}~.
\eea

To summarize: the unpaired $B$ modes are $|\mathcal{P}_+/\Omega_+|^{\qq/2}$ times a holomorphic function on $\C^2$ with respect to the $I_+$ complex 
structure, while the unpaired $\phi$ modes  are $|\mathcal{P}_-/\Omega_-|^{-\qq/2}$ times a holomorphic function on $\C^2$ with respect to the $I_-$ complex structure. 
Here $\qq=r-2$ for $B$, while $\qq=r$ for $\phi$, where $r$ is the R-charge of the matter multiplet.

As discussed in section \ref{sec:unpaired}, the contributions of these modes to the supersymmetric Casimir energy is determined by their eigenvalues 
under $\ii \hL_{\frac{1}{2}\xi}$, where recall that acting on scalars
\bea
\ii \hL_{\frac{1}{2}\xi} &=& \frac{\ii}{2}\mathcal{L}_\xi + \frac{\qq}{2}\gamma~, 
\eea
where 
\bea
\gamma &\equiv & -\mathcal{L}_{\frac{1}{2}\xi}\omega \ = \ \frac{1}{2}(|b_1|+|b_2|)~.
\eea
The eigenvalues are then easily computed:
\bea\label{Bandphieigenvalues}
\ii \hL_{\frac{1}{2}\xi} B_{n_1,n_2} &=& \frac{1}{2}\left[-n_1|b_1| - n_2|b_2| + \qq\gamma\right]B_{n_1,n_2} \ = \ \lambda^B_{n_1,n_2}B_{n_1,n_2}~, \nn\\
\ii \hL_{\frac{1}{2}\xi} \phi_{n_1,n_2} &=& \frac{1}{2}\left[n_1|b_1| + n_2|b_2| +\qq\gamma\right]\phi_{n_1,n_2}  \ = \ \lambda^\phi_{n_1,n_2}\phi_{n_1,n_2}~,
\eea
where we have used that the Reeb vector is given by (\ref{ReebHopf}), and hence
\bea
\mathcal{L}_\xi z_i^\pm &=& \pm \ii |b_i| z_i^\pm~, \qquad i = 1,2~.
\eea

We may now further reinterpret the eigenvalues $\lambda^B$, $\lambda^\phi$, using our earlier description of the holomorphic volume 
forms $\Omega_\pm$. Recall that $A$ is a connection on $\mathcal{K}_+^{-1/2}$, so that 
$\bar{\partial}^+_A$ acts on sections of $\mathcal{K}^{-\qq/2}_+$. In the case at hand 
$\mathcal{K}_+\cong \Lambda^{2,0}_+$ is a trivial bundle over $\C^2\setminus\{0\}$, 
but the holomorphic section $\Omega_+$ of $\mathcal{K}_+$ leads to a canonical lifting of the $U(1)\times U(1)$ 
action, with generators $\partial_{\psi_1}$, $\partial_{\psi_2}$, to the fibre. Specifically, 
since $\Omega_+$ satisfies $\mathcal{L}_{\partial_{\psi_i}}\Omega_+=\ii \Omega_+$, 
$i=1,2$, the generators $(q_1,q_2)\in U(1)\times U(1)$ acting on $\C^2=\C\oplus\C$
as $(z_1^+,z_2^+)\rightarrow (q_1z_1^+,q_2z_2^+)$ 
act as multiplication by $q_1q_2$ on $\Omega_+$. With this understanding, 
the eigenvalue $\lambda^B$ is the eigenvalue of the \emph{ordinary} Lie derivative 
$\ii\mathcal{L}_{\frac{1}{2}\xi}$ acting on holomorphic sections of $\mathcal{K}^{-\qq/2}_+$. 
Here the action on the fibre contributes precisely $-\frac{\qq}{2}(-\gamma)=\frac{\qq}{2}\gamma$ to 
$\ii\mathcal{L}_{\frac{1}{2}\xi}$, since $\ii\mathcal{L}_{\frac{1}{2}\xi}\Omega_+=-\gamma\Omega_+$. 

A similar reasoning applies to the $\phi$ modes. Here $-A$ is a connection on $\mathcal{K}_-^{-1/2}$, 
so that $\bar{\partial}^-_A$ acts on sections of $\mathcal{K}^{\qq/2}_-$. Again the canonical 
bundle is trivial, but the action of $(q_1,q_2)\in U(1)\times U(1)$ above on the fibre is now 
$(q_1q_2)^{-1}$. This follows from the relative minus signs in the phases in (\ref{zplus}),
(\ref{zminus}). The action on the fibre then again contributes precisely $\frac{\qq}{2}\gamma$ to $\ii\mathcal{L}_{\frac{1}{2}\xi}$,
since now $\ii\mathcal{L}_{\frac{1}{2}\xi}\Omega_-=\gamma\Omega_-$ . 

Notice that with these definitions $\mathcal{K}_+\cong (\mathcal{K}_-)^{-1}$ as equivariant holomorphic line bundles under $U(1)\times U(1)$.

\subsection{The character}

The supersymmetric Casimir energy is (formally, before regularization) 
\bea\label{Esusy}
\Esusym &=& \sum_{n_1,n_2\in \Z_{\geq 0}}\lambda^B_{n_1,n_2} +  \sum_{n_1,n_2\in \Z_{\geq 0}}\lambda^\phi_{n_1,n_2}~,
\eea
where the eigenvalues are those on the right hand side of (\ref{Bandphieigenvalues}). Here we have introduced the 
superscript ``matter'' to emphasize that in what follows we focus 
on the contribution of a single weight $\rho$ in a weight space decomposition 
of the chiral matter representation $\mathcal{R}$.
We have seen that the eigenvalues in (\ref{Esusy}) are precisely Reeb charges, under 
$\ii\mathcal{L}_{\frac{1}{2}\xi}$, of holomorphic sections of $\mathcal{K}_+^{-\qq/2}$
and $\mathcal{K}_-^{\qq/2}$, respectively, where $\qq=r-2$ for the $B$ modes and 
$\qq=r$ for the $\phi$ modes. Thus it is natural at this point to introduce 
the index-character of \cite{Martelli:2006yb} that counts such holomorphic 
sections according to their $U(1)\times U(1)$ charges. We take the $U(1)\times U(1)$ 
generators to be $(q_1,q_2)$, which act as 
\bea
(z_1^\pm,z_2^\pm)&\rightarrow & (q_1^{\pm 1}z_1^\pm,q_2^{\pm 1}z_2^\pm)~.
\eea
For the $B$ modes we have the associated index-character
\bea\label{LB}
\Lef(\bar{\partial}_{\mathcal{K}_+^{-\qq/2}},(q_1,q_2)) &=& \sum_{n_1,n_2\geq 0} (q_1q_2)^{-\qq/2}\cdot q_1^{n_1}q_2^{n_2}~.
\eea
The left hand side is defined as the trace of the action of $(q_1,q_2)$ on the zero modes of the operator
$\bar{\partial}_{\mathcal{K}_+^{-\qq/2}}$.
The right hand side of (\ref{LB}) is a divergent series for $|q_1|=|q_2|=1$, but by analytically continuing to 
$|q_1|,|q_2|<1$ the series converges to give
\bea
\Lef(\bar{\partial}_{\mathcal{K}_+^{-\qq/2}},(q_1,q_2)) &=& \frac{(q_1q_2)^{-\qq/2}}{(1-q_1)(1-q_2)}~.
\label{bchara}
\eea
This then effectively regularizes the eigenvalue sum. Indeed, 
  setting $q_1=\ex^{t|b_1|}$, $q_2=\ex^{t|b_2|}$ and formally 
expanding (\ref{LB}) in a Taylor series around $t=0$, the 
coefficient of $-t$ is precisely
$2\lambda_{n_1,n_2}^B=-n_1|b_1|-n_2|b_2|+\qq\gamma$. Recalling that  the $B$ modes have $\qq=r-2$, we hence see 
that according to this ``character regularization'' their contribution to the 
supersymmetric Casimir energy is
\bea\label{EB}
E_{\mathrm{susy}}^B &= & \frac{1}{2} \left[\left.\frac{(q_1q_2)^{-(r-2)/2}}{(1-q_1)(1-q_2)}\right|_{q_1\, = \, \ex^{t|b_1|},\, q_2\, = \, \ex^{t|b_2|}}\right]_{\mathrm{coefficient}\,  \mathrm{of}\,  -t}\nn\\
&=& \frac{1}{96|b_1||b_2|}(|b_1|+|b_2|)(r-1)\left[(|b_1|+|b_2|)^2(r-1)^2 -(b_1^2+b_2^2)\right]\nn\\
&=& \frac{1}{2}\cdot \left.\frac{4u^3-(b_1^2+b_2^2)u}{24|b_1||b_2|}\right|_{u \, = \, (r-1)\gamma}~,
\eea
where the second equality is by a simple direct computation. This is indeed the correct contribution 
of the unpaired $B$ modes to the supersymmetric Casimir energy! 

The $\phi$ modes work similarly. The relevant character is now
\bea\label{Lpartial}
\Lef(\bar{\partial}_{\mathcal{K}_-^{\qq/2}},(q_1,q_2)) &=& \sum_{n_1,n_2\geq 0} (q_1q_2)^{-\qq/2}\cdot q_1^{-n_1}q_2^{-n_2}~.
\eea
Summing the series for $|q_1|,|q_2|>1$ we obtain
\bea\label{phich}
\Lef(\bar{\partial}_{\mathcal{K}_-^{\qq/2}},(q_1,q_2)) &=&  
\frac{(q_1q_2)^{-\qq/2}}{(1-q_1^{-1})(1-q_2^{-1})} \ = \ \frac{(q_1q_2)^{-(\qq-2)/2}}{(1-q_1)(1-q_2)}~.
\eea
Recalling that 
$\phi$ has R-charge $r$, we see that their contribution is also precisely the right hand side of the 
first line of (\ref{EB}). Thus they contribute equally to the supersymmetric Casimir energy, as expected, 
$E_{\mathrm{susy}}^\phi=E_{\mathrm{susy}}^B$.

\subsection{Zeta function versus heat kernel regularization}

At first sight the result just obtained is somewhat remarkable, 
because we regularized the eigenvalue sum (\ref{Esusy})
using the index-character (via analytic continuation to a simple geometric series), 
while in previous work the sum in (\ref{Esusy}) is regularized using the Barnes 
double zeta function. The two regularization schemes lead to the same result. 

This may be explained as follows. In order to regularize each sum in 
 (\ref{Esusy}) in a supersymmetric fashion one should 
replace\footnote{Below $n$ denotes a multi-index.}
\be
\sum_n \lambda_n \ \to  \ \sum_n \lambda_n \, f(\lambda_n , t) ~,
\ee
with $f(x,t)$ a function chosen so that the sum converges. Requiring that $f(x,0)=1$, the value of the  regularized sum is given by the \emph{finite part}
in the limit that the parameter $t\to 0$. Indeed, supersymmetric counterterms exist that may be added to remove divergences appearing as  poles in
 $t^{-2}$ and $t^{-1}$. However, the fact that  finite supersymmetric counterterms do not exist   \cite{Assel:2014tba} implies that the finite part is  unambiguous, and therefore  independent of the
 details of the regularization.  There are two natural choices. Picking  $f(\lambda_n , t)= \lambda_n^{-t}$ leads to the spectral zeta function regularization, while 
 the choice  $f(\lambda_n , t)= \me^{-t\lambda_n}$ leads to the heat kernel regularization, which as we shall see 
 is the ``character regularization'' we have used above. It is well known that these two are related to each other via the Mellin transform. 
 
In the case of interest the sums in (\ref{Esusy}) were regularized in  \cite{Assel:2015nca} using 
the Barnes double zeta function, defined as
\bea\label{zeta2}
\zeta_2(t;|b_1|,|b_2|,x) & \equiv & \sum_{n_1,n_2\in \Z_{\geq 0}} (|b_1|n_1+|b_2|n_2+x)^{-t}~,
\eea
where $x=r\gamma$ for the physical case of interest. Here we have focused on the $\phi$ modes.
The sum in (\ref{zeta2}) converges for $\mathrm{Re}\, t>1$ and one analytically continues  to $t=-1$ obtaining  \cite{spreafico}
\bea\label{cubic}
\Esusym &=& 
 \frac{u^3}{6|b_1||b_2|} - \frac{(b_1^2+b_2^2)u}{24|b_1||b_2|}~,
\eea
where we have defined $u=(r-1)\gamma = x - \gamma$.  Note that 
\bea
\frac{1}{2}\zeta_2\left(-1;|b_1|,|b_2|,u+\gamma\right)  & =  &  -   \frac{1}{2}\zeta_2\left(-1;|b_1|,|b_2|,-u+\gamma\right)~,
\label{Bsamephi}
\eea
so that the contributions to $\Esusym$ of the modes $\phi$ and $B$ are indeed identical. 

Alternatively, in the heat kernel regularization we are led to consider 
\bea
S(t; |b_1|,|b_2|,x) & \equiv &   \sum_{n_1,n_2\in \Z_{\geq 0}} \me^{-t (|b_1|n_1+|b_2|n_2+x)}~,
\eea
and we extract $\Esusym$ from the coefficient of $-t$ in a series around $t=0$. This is 
precisely the character regularization we introduced above. Concretely,
\bea\label{heatcasimir}
S(t; |b_1|,|b_2|,x)  & = & \frac{\me^{-t x}}{(1-\me^{-t|b_1|})(1-\me^{-t|b_2|})}\nonumber\\
& = & \left.\frac{(q_1q_2)^{-r/2}}{(1-q_1^{-1})(1-q_2^{-1})}\right|_{q_1\, = \, \ex^{t|b_1|},\, q_2\, = \, \ex^{t|b_2|}}~,
\eea
where recall that $x=r\gamma$, and in the second line we have precisely the character (\ref{phich}) for the $\phi$ modes.

\subsection{Rewriting as a Dirac character}

In the above discussion we saw that both the $B$ and $\phi$ unpaired modes lead to the same contribution to the supersymmetric 
Casimir energy. However, the discussion is not quite symmetric because $B$ has R-charge $\qq=r-2$, while 
$\phi$ has R-charge $\qq=r$. One can put these on the same footing, with overall R-charge $r-1$, by effectively further
twisting the $\bar\partial$ operators, thus viewing them as (part of) a Dirac operator.

Let us begin with the $\phi$ zero modes. The relevant operator is
\bea
\bar{\partial}_{\mathcal{K}_-^{r/2}} & \cong & \bar{\partial}_{\mathcal{K}_-^{1/2}\otimes\, \mathcal{L}_-}~,
\eea
where we have defined
\bea
\mathcal{L}_- &\equiv & \mathcal{K}_-^{(r-1)/2}~.
\eea
Let us denote the weight on $\mathcal{L}_-$ as $\lambda=(q_1q_2)^{-(r-1)/2}$. Then the relevant character is
\bea
\Lef(\bar{\partial}_{\mathcal{K}_-^{r/2}},(q_1,q_2)) &=& \frac{(q_1q_2)^{-1/2}}{(1-q_1^{-1})(1-q_2^{-1})} \lambda \ = \ 
\frac{(q_1q_2)^{1/2}}{(1-q_1)(1-q_2)}\lambda~,
\eea
where the $(q_1q_2)^{-1/2}$ in the numerator comes from the twisting by $\mathcal{K}_-^{1/2}$. 

Similarly for the $B$ zero modes the operator is
\bea
\bar\partial_{\mathcal{K}_+^{-(r-2)/2}} & \cong & \partial_{\mathcal{K}_+^{1/2}\otimes \, \mathcal{L}_+}~,
\eea
where
\bea
\mathcal{L}_+ & \equiv & \mathcal{K}_+^{-(r-1)/2}~.
\eea
Notice that the weight on $\mathcal{L}_+$ is also $\lambda=(q_1q_2)^{-(r-1)/2}$, and indeed 
$\mathcal{L}_+\cong \mathcal{L}_-$. Thus the relevant character is
\bea
\Lef(\bar\partial_{\mathcal{K}_+^{-(r-2)/2}},(q_1,q_2)) &=& \frac{(q_1q_2)^{1/2}}{(1-q_1)(1-q_2)}\lambda~.
\eea
This makes manifest that the two modes have the \emph{same} character. In both cases the operator is 
$\bar\partial^\pm$ twisted by $\mathcal{K}_\pm^{1/2}\otimes \mathcal{L}_\pm$, which may be viewed as 
part of a Dirac-type operator twisted by $\mathcal{L}_\pm$. From this point of view, the explicit 
$(q_1q_2)^{1/2}$ factors come from the fact that the modes transform as spinors under the $U(1)\times U(1)$ action.

We may thus define
\bea\label{LDirac}
\Lef(\mathrm{Dirac},(q_1,q_2,\lambda)) &\equiv & \frac{(q_1q_2)^{1/2}}{(1-q_1)(1-q_2)}\lambda~.
\eea
Setting $q_1=\ex^{t|b_1|}$, $q_2=\ex^{t|b_2|}$, $\lambda=\ex^{-tu}$, we may expand in a Laurent series around $t=0$ 
as above:
\bea\label{indexcasimir}
\Lef(\mathrm{Dirac},(q_1,q_2,\lambda))  & = & 
\frac{1}{|b_1||b_2| t^2} - \frac{u}{|b_1||b_2| t} + \Big(\frac{u^2}{2! |b_1||b_2|}-\frac{b_1^2+b_2^2}{24 |b_1||b_2|}\Big)\nn\\
&&
-\Big(\frac{u^3}{ 3! |b_1||b_2|}-\frac{(b_1^2+b_2^2)u}{24|b_1||b_2|}\Big)t \\
&&+ \Big(\frac{u^4}{4! |b_1||b_2|}-\frac{(b_1^2+b_2^2)u^2}{24\cdot 2!|b_1||b_2|}+\frac{7(b_1^2+b_2^2)-4b_1^2b_2^2}{5760|b_1||b_2|}\Big)t^2 + O(t^3)~.\nn
\eea
We immediately see that the divergent ``index'', which is given by setting $t=0$, arises as a second order pole, while  
the coefficient of the linear term in $-t$ precisely reproduces the regularized supersymmetric 
Casimir energy (setting $u$ to its physical value of $u=(r-1)\gamma$). 
Of course this is simply equivalent to the computation in (\ref{EB}), although now the equal contribution 
of the $B$ and $\phi$ modes is manifest.

The appearance of  the $A$-roof class in the expansion (\ref{indexcasimir}) is explained by the following identity:
\bea\label{fpt}
\frac{\hat{A}(\ii\theta_1,\ii\theta_2)}{\chi(\ii\theta_1,\ii\theta_2)} &=& \frac{1}{(q_1^{1/2}-q_1^{-1/2})(q_2^{1/2}-q_2^{-1/2})} \ = \ 
\frac{(q_1q_2)^{1/2}}{(1-q_1)(1-q_2)}~,
\eea
where $q_1=\ex^{\ii\theta_1}$, $q_2=\ex^{\ii\theta_2}$.
Here the numerator on the left hand side is the $A$-roof class, which in general is defined as 
\bea\label{Aroof}
\hat{A}(x_1,\ldots,x_n) &=& \prod_{i=1}^n\frac{ x_i}{\ex^{x_i/2}-\ex^{-x_i/2}}\ = \  \prod_{i=1}^n\frac{ x_i}{2\sinh x_i/2}~,
\eea
while the denominator is the Euler class
\bea\label{Euler}
\chi(x_1,\ldots,x_n) &=& \prod_{i=1}^nx_i~.
\eea
In the usual index theorem the $x_i$ would be the first Chern classes of the 
line bundles that arise on application of the splitting principle. In the equivariant setting these are replaced by $x_i+\ii \theta_i$, 
where the group action on the complex line fibre is multiplication by $\ex^{\ii\theta_i}$. 
The Euler class cancels against the numerator of (\ref{Aroof}), which leads to the first equality in (\ref{fpt}).
The $A$-roof class may be expanded as
\bea
\hat{A} &=& 1 - \frac{1}{24}p_1 + \frac{1}{5760}(7p_1^2-4p_2 )+\cdots~,
\eea
where the Pontryagin classes $p_I$ are the $I$th elementary symmetric functions in the 
$x_i^2$. Thus in particular for complex dimension $n=2$ we have $p_1=x_1^2+x_2^2$, 
$p_2=x_1^2x_2^2$. These comments of course explain the structure of the right hand side 
of (\ref{indexcasimir}). Analytically continuing $q_1=\ex^{t|b_1|}, q_2=\ex^{t|b_2|}$ amounts to sending $\ii\theta_i\rightarrow t|b_i|$ above.
Then (\ref{indexcasimir}) may be rewritten as
\bea
\Lef(\mathrm{Dirac},(q_1,q_2,\lambda))& = & \frac{\ex^{-tu}}{4\sinh (t|b_1|/2) \sinh (t|b_2|/2)}\\
&=&
\frac{1}{|b_1||b_2| t^2}\left(1- \frac{b_1^2+b_2^2}{24}t^2 + \frac{7(b_1^2+b_2^2)^2-4b_1^2b_2^2}{5760}t^4+\cdots\right)\ex^{-tu}~.\nn
\eea
The middle term in brackets is the contribution from the $A$-roof class.
This of course explains the observation in \cite{Bobev:2015kza} that the supersymmetric 
Casimir energy on the primary Hopf surface is obtained  (formally) by an equivariant integral on $\R^4$ associated
to the Dirac operator. This arises naturally in the way we have formulated the problem. Here
 the supersymmetric Casimir energy is the coefficient of $-t$ in an expansion of the index-character 
of the Dirac operator, where the latter is regularized by analytically continuing 
a divergent geometric series into its domain of convergence. 
Mathematically, this arises as a heat kernel regularization, as opposed to a 
(Barnes) zeta function regularization.

\section{Secondary Hopf surfaces and generalizations}
\label{moregeneral:sec}

\subsection{Lens spaces}
\label{sec:lens}

The simplest way to generalize the primary Hopf surfaces studied in the previous 
section is to take a $\Gamma\cong \Z_p$ quotient. These secondary 
Hopf surfaces were described at the beginning of section \ref{secondaryHopfgeometry}. 
With respect to either complex structure $I_\pm$ the $\Z_p$ action is generated by 
$(z_1,z_2)\rightarrow (\ex^{2\pi \ii/p}z_1,\ex^{-2\pi \ii/p}z_2)$, where $z_i=z_i^\pm$, $i=1,2$. This action preserves the 
Killing spinors $\zeta_\pm$, and hence in particular the function $s$ and holomorphic volume forms on $\C^2$. 
The quotient $M_3=S^3/\Z_p=L(p,1)$ is then a Lens space.

Since $\pi_1(M_3)\cong \Z_p$, the space $\R\times M_3$ now supports non-trivial flat connections. 
As discussed in section \ref{sec:Casimir}, the localized partition function on $S^1\times M_3$ splits 
into associated topological sectors, which are summed over. In the Hamiltonian approach, each such sector leads to
a distinct supersymmetric quantum mechanics on $\R$. 
Following the end of section \ref{sec:flat}, here we consider 
a $U(N)$ gauge theory with matter in a representation $\mathcal{R}$ 
in a weight space decomposition. The modes $B=B_\rho$, $\phi=\phi_\rho$ 
then become sections of $\mathcal{K}_+^{-\qq/2}\otimes L$ and 
$\mathcal{K}_-^{\qq/2}\otimes L$, respectively, where 
 the line bundle $L$ over $\R\times L(p,1)$ 
has first Chern class $c_1(L)\equiv \rho(\mathbf{m})$ mod $p$. 
In the Hamiltonian approach, and for fixed topological sector $\mathbf{m}$, 
we thus want to compute a \emph{twisted character}, which 
counts holomorphic sections of $\mathcal{K}_+^{-\qq/2}\otimes L$ and 
$\mathcal{K}_-^{\qq/2}\otimes L$ according to their $U(1)\times U(1)$ charges 
 (where as usual $\qq=r-2$ for $B$ and $\qq=r$ for $\phi$).

Recall that holomorphic functions on $\C^2$ are counted by 
\bea\label{charC2}
\Lef(\bar{\partial},(q_1,q_2),\C^2) &=& \frac{1}{(1-q_1)(1-q_2)}~.
\eea
The Dirac index-character (\ref{LDirac}) is constructed from this by multiplying 
by $(q_1q_2)^{1/2}\lambda$, which takes account of the lifting 
of the $U(1)\times U(1)$ action to $\mathcal{K}_\pm^{\mp \qq/2}$. 
More generally, holomorphic sections of $L$ over $\C^2/\Z_p$, where $c_1(L) \equiv \holonomy$ mod $p$, are 
counted by the twisted character
\bea\label{twistedchar}
\Lef(\bar{\partial}_L,(q_1,q_2),\C^2/\Z_p) &=& \frac{q_1^\holonomy(1-(q_1q_2)^{p-\holonomy})+ q_2^{p-\holonomy}(1-(q_1q_2)^\holonomy)}{(1-(q_1q_2))(1-q_1^p)(1-q_2^p)}~.
\eea
Here $\nu$ is understood to lie in the range $0\leq \nu<p$, and 
 as usual one expands the denominator in a geometric series, for $|q_1|,|q_2|<1$. 
Perhaps the simplest way to derive (\ref{twistedchar}) is via an appropriate 
projection of (\ref{charC2}).
Recall that the $\Z_p$ action on $\C^2$ is generated by $(z_1,z_2)\rightarrow (\omega_pz_1,\omega_p^{-1}z_2)$, 
where $\omega_p\equiv\ex^{2\pi\ii/p}$.
The twisted character is then
\bea
\Lef(\bar{\partial}_L,(q_1,q_2),\C^2/\Z_p) &=& \frac{1}{p}\sum_{j=0}^{p-1} \frac{\omega_p^{-j\holonomy}}{(1-\omega_p^j q_1)(1-\omega_p^{-j}q_2)}~.
\eea
One easily verifies that this may be simplified to give (\ref{twistedchar}).
For zero twist, meaning $\holonomy=0$, 
we are simply counting holomorphic functions on $\C^2/\Z_p$, and (\ref{twistedchar}) reads
\bea\label{characterLens}
\Lef(\bar{\partial},(q_1,q_2),\C^2/\Z_p) &=& \frac{(1-(q_1q_2)^p)}{(1-(q_1q_2))(1-q_1^p)(1-q_2^p)} \nn\\
&= & \frac{1+q_1q_2+(q_1q_2)^2 + \cdots + (q_1q_2)^{p-1}}{(1-q_1^p)(1-q_2^p)}~.
\eea
This is the index-character of an $A_{p-1}=\C^2/\Z_p$ singularity.

Thus the contribution of a matter field,
for weight $\rho$ and fixed flat connection $\mathbf{m}$, leads to
a supersymmetric Casimir energy (in the sector $\iflat=\mathbf{m}\in\Mflat$) given by the character
\bea
\frac{(q_1q_2)^{1/2}\left[q_1^\holonomy(1-(q_1q_2)^{p-\holonomy})+ q_2^{p-\holonomy}(1-(q_1q_2)^\holonomy)\right]}{(1-(q_1q_2))(1-q_1^p)(1-q_2^p)}\lambda~.
\eea
As in section \ref{sec:primary}, the Casimir energy is obtained by
substituting $q_1=\ex^{t|b_1|}$, $q_2=\ex^{t|b_2|}$, $\lambda=\ex^{-tu}$, and extracting the coefficient of $-t$ in a Laurent series around $t=0$. This is easily done, and we find
\bea\label{esusylens}
\Esusymm &=& \frac{1}{24|b_1||b_2|p}\big[4u^3 - (b_1^2+b_2^2 - 2|b_1||b_2|(p^2 - 6\holonomy p+6\holonomy^2-1))u \nn\\
&&+ 2|b_1||b_2|(|b_1|-|b_2|)\holonomy(\holonomy-p)(2\holonomy-p) \big]~,
\eea
where $\holonomy=\widehat{\rho(\mathbf{m})}$. Here the hat indicates that $\nu$ is understood to lie in the range $0\leq \nu<p$, 
and thus $\rho(\mathbf{m})\in \Z$ should be reduced mod $p$ to also lie in this range. Recall that we fixed the convention 
that $0\leq m_i<p$, and ordered $m_1\leq \cdots \leq m_N$.
As usual we should also put $u=(r-1)\gamma$, where $\gamma=(|b_1|+|b_2|)/2$. 
This is the contribution from the weight $\rho$; one should of course then sum over weights to get the total 
contribution of the matter field, in the sector $\mathbf{m}$. 

The partition function on $S^1\times L(p,1)$
 has been computed in \cite{Nishioka:2014zpa}, and fixing the sector $\mathbf{m}$ one can 
check that  indeed
\bea
\Esusymm  &=& -\lim_{\beta\rightarrow \infty} \frac{\diff}{\diff\beta} \log Z^{\mathrm{matter}}_{\mathbf{m}}~.
\eea
See equations (5.32)--(5.34) of \cite{Nishioka:2014zpa}. 
Thus the Hamiltonian approach does indeed correctly reproduce the supersymmetric Casimir energy, defined 
in terms of the partition function, for each topological sector.

\subsection{Fixed point formula}\label{sec:fp}

In \cite{Martelli:2006yb} it was explained that the index-character may be computed for a general 
isolated singularity by first resolving the singularity, and using a fixed point formula. 
In the case at hand $\C^2/\Z_p=A_{p-1}$ is well-known to admit a crepant resolution, meaning that 
the holomorphic $(2,0)$-form extends smoothly to 
the resolved space,
by blowing up $p-1$ two-spheres.  The action of  $U(1)\times U(1)$ on $\C^2/\Z_p$ extends to the resolution, 
which is hence \emph{toric}, with $p$ isolated fixed points. 
Each such fixed point is of course locally modelled by $\C^2$, and 
the general formula in \cite{Martelli:2006yb} expresses the index-character 
of $\C^2/\Z_p=A_{p-1}$ in terms of a sum of the  index-characters for $\C^2$, for each fixed point. 
Labelling the fixed points by $j=0,\ldots,p-1$, explicitly we have
\bea\label{characterLensfp}
\Lef(\bar{\partial},(q_1,q_2),\C^2/\Z_p) &=& \sum_{j=0}^{p-1} \frac{1}{(1-q_1^{u_1^{(j)}}q_2^{u_2^{(j)}})(1-q_1^{v_1^{(j)}}q_2^{v_2^{(j)}})}~.
\eea
Here the action of $U(1)\times U(1)$ on each fixed origin of $\C^2$ is specified by the 
two vectors $\mathbf{u}^{(j)}=(u_1^{(j)},u_2^{(j)}), \mathbf{v}^{(j)}=(v_1^{(j)},v_2^{(j)})\in\Z^2$ as
\bea
(z_1,z_2) & \rightarrow & (q_1^{u_1^{(j)}}q_2^{u_2^{(j)}}z_1,q_1^{v_1^{(j)}}q_2^{v_2^{(j)}}z_2)~.
\eea
One finds (for example using toric geometry methods) that
\bea
\mathbf{u}^{(j)} &=& (p-j,-j)~, \qquad \mathbf{v}^{(j)} \ = \  (-p+j+1,j+1)~,
\eea
and (\ref{characterLensfp}) reads
\bea
\Lef(\bar{\partial},(q_1,q_2),\C^2/\Z_p) &=& \sum_{j=0}^{p-1} \frac{1}{(1- q_1^{p-j}q_2^{-j})(1- q_1^{-p+j+1}q_2^{j+1})}~,
\eea
which one can verify agrees with (\ref{characterLens}).

Let us define the matter contribution to the supersymmetric Casimir energy for $S^3$ as
\bea
\Esusym[S^3; b_1,b_2] &=& \frac{4u^3 - (b_1^2+b_2^2)u}{24b_1b_2}~.
\eea
Then (\ref{characterLensfp}) leads to the following fixed point formula for the 
Casimir for $S^1\times L(p,1)$ (with trivial flat connection):
\bea
\Esusym[L(p,1);b_1,b_2] & =& \sum_{j=0}^{p-1}\Esusym[S^3;b_1^{(j)},b_2^{(j)}]\nn\\
&=& \frac{4u^3-[(|b_1|+|b_2|)^2-2|b_1||b_2|p^2]u}{24|b_1||b_2|p}~.
\eea
Here we have defined
\bea\label{Reebk}
b_1^{(j)} &\equiv & p|b_1|-j(|b_1|+|b_2|)~, \qquad b_2^{(j)} \ \equiv \ -p|b_1|+(j+1)(|b_1|+|b_2|)~.
\eea
In fact  $(b_1^{(j)},b_2^{(j)})$, $j=0,\ldots,p-1$,
are precisely the Reeb weights at the $p$ fixed points. 
In this precise sense, we may write the supersymmetric Casimir energy 
for the secondary Hopf surface $(S^1\times S^3)/\Z_p$ 
as the sum of $p$ Casimir energies for primary Hopf surfaces 
$S^1\times S^3$, where each fixed point contribution has a 
different complex structure, determined by (\ref{Reebk}). 
This data is in turn determined by the equivariant geometry of the 
resolved space.

\subsection{More general $M_3$}

In section \ref{secondaryHopfgeometry} we discussed more general 
classes of secondary Hopf surfaces, realised as $\Gamma=\Gamma_{ADE}\subset SU(2)$ 
quotients of primary Hopf surfaces. The A series is precisely the Lens space case
discussed in the previous subsection, while the D and E series result in non-Abelian 
fundamental groups. The formalism we have described gives a prescription 
for computing the supersymmetric Casimir energy $\Esusy$ (or at least the matter 
contribution $\Esusym$) for such 
backgrounds. 
One first needs to classify the inequivalent flat $G$-connections on $M_3=S^3/\Gamma$, via their corresponding homomorphisms
$\varrho:\Gamma\rightarrow G$. A given matter representation $\mathcal{R}$ of $G$ then gives 
a corresponding flat $\mathcal{R}$-connection, from which one constructs the 
matter bundle (\ref{flatmatter}). For each such flat connection one then needs 
to compute the index-character of this bundle, namely one counts holomorphic sections 
via their Reeb charges. The supersymmetric Casimir energy, in this topological sector, 
is then obtained as a limit of this index-character.

\subsubsection{Poincar\'e Hopf surface}\label{sec:Poincare}

In practice, one thus first needs to understand the representation theory 
of the relevant non-Abelian groups, before one can compute the associated 
index-characters. An interesting but simple example 
is provided by the exceptional group $\Gamma=\Gamma_{E_8}$: 
this is the binary icosahedral group, which has order 120. 
The quotient $M_3=S^3/\Gamma$ is the famous Poincar\'e sphere, 
which has the homology groups of $S^3$, despite the very large 
fundamental group. This follows since $\Gamma_{E_8}$ is 
equal to its commutator subgroup, and hence its Abelianization (which 
equals $H_1(M_3,\Z)$) is trivial. Related to this fact is that 
consequently any homomorphism into an \emph{Abelian} group 
is necessarily trivial. This is easy to see: since 
any group element $g\in \Gamma$ may be written as $g=hvh^{-1}v^{-1}$, 
then for any homomorphism $\varrho:\Gamma\rightarrow G$ we have
$\varrho(g)=\varrho(h)\varrho(v)\varrho(h)^{-1}\varrho(v)^{-1}=$ identity, 
where in the last step we used that $G$ is Abelian. This shows that, for 
example, any flat $U(1)$ connection over the Poincar\'e sphere is necessarily trivial. 
Because of this, to compute the supersymmetric Casimir energy we need only 
the index-character of $\C^2/\Gamma$. But this is easily computed 
by realizing the latter as a homogeneous hypersurface singularity
\bea
\C^2/\Gamma_{E_8} &\cong & \{f_{E_8} \ \equiv \  Z_1^3+Z_2^5+Z_3^2 \ = \ 0\}\subset \C^3~.
\eea
Here the polynomial $f_{E_8}$ has degree $d=30$ under the weighted $\C^*$ action on $\C^3$ with weights
 $(w_1,w_2,w_3)=(10,6,15)$. From the general formula in \cite{Gauntlett:2006vf} we thus compute 
 the index-character
 \bea
 \Lef(\bar{\partial},q,\C^2/\Gamma_{E_8}) \ =  \ \frac{1-q^{30}}{(1-q^{6})(1-q^{10})(1-q^{15})} \ = \ 1+q^6+q^{10}+q^{12}+q^{15}+\ldots~.
 \eea
Here $q\in\C^*$ acts diagonally on $\C^2/\Gamma_{E_8}$ as $(z_1,z_2)\rightarrow (q^{1/2}z_1,q^{1/2}z_2)$. 
Notice that the centre of $\Gamma_{E_8}$ is $\Z_2$, which acts as multiplication on $(z_1,z_2)$ by $-1$. 
The holomorphic $(2,0)$-form  thus has weight $q$ under the $\C^*$ action, and the 
supersymmetric Casimir energy for an Abelian gauge theory 
on the ``Poincar\'e Hopf surface''  is
\bea
\Esusym &=& \left[\left. q^{1/2}\lambda\cdot \frac{1-q^{30}}{(1-q^{10})(1-q^6)(1-q^{15})}\right|_{q\, =\, \ex^{t|b|}, \, \lambda \, =\,  \ex^{-tu}}\right]_{\mathrm{coefficient}\,  \mathrm{of}\,  -t}\nn\\
&=& \frac{4u^3+539b^2u}{720b^2}~.
\eea
The Reeb vector field acting on $(z_1=|z_1|\ex^{\ii\psi_1},z_2=|z_2|\ex^{\ii\psi_2})$ is 
\bea
\xi &=& \partial_\psi \ = \ \frac{|b|}{2}(\partial_{\psi_1}+\partial_{\psi_2})~,
\eea
while for a matter multiplet of R-charge $r$ we have $u=(r-1)|b|/2$.

\subsubsection{Homogeneous hypersurface singularities}

For an Abelian gauge theory on the Poincar\'e Hopf surface just discussed, any flat connection 
over $S^3/\Gamma_{E_8}$ is trivial, and thus the index-character that counts 
holomorphic functions on $\C^2/\Gamma_{E_8}$ is sufficient to compute the supersymmetric Casimir energy. 
However, more generally we may easily extend the above discussion to compute 
$\Esusy$ for $\Z$ quotients of homogeneous hypersurface singularities 
 in the sector with trivial flat connection. These are compact complex surfaces 
 of the form $M_4=S^1\times M_3$, where $M_3$ is the link of the 
 singularity.
 
Consider a general   weighted homogeneous 
hypersurface singularity in $\C^3$. Here the  weighted $\C^*$ action on $\C^3$ is
$(Z_1,Z_2,Z_3)\rightarrow (\Nlambda^{w_1}Z_1,\Nlambda^{w_2}Z_2,\Nlambda^{w_3}Z_3)$,
 where $w_i\in\mathbb{N}$ are the weights, $i=1,2,3$, and $\Nlambda\in\C^*$. 
 The hypersurface is the zero set $X\equiv \{f=0\}\subset\C^3$ of a weighted homogeneous polynomial 
 $f=f(Z_1,Z_2,Z_3)$, where
 \bea
 f(\Nlambda Z_1,\Nlambda Z_2,\Nlambda Z_3) &=& \Nlambda^d f(Z_1,Z_2,Z_3)~,
 \eea
 which defines the degree  $d\in\mathbb{N}$. We assume that $f$ is such that 
  $X\setminus\{o\}\cong \R\times M_3$ is smooth, where $o$ is the origin $Z_1=Z_2=Z_3=0$.
   The associated compact 
  complex surface is obtained as a free $\Z$ quotient of $X\setminus\{0\}$, where 
   $\Z\subset\C^*$ is embedded as $n\rightarrow \Nlambda^n$
for some fixed $\Nlambda>1$. The Reeb vector field action is quasi-regular, generated by 
$\Nlambda\in U(1)\subset \C^*$, and the quotient $\Sigma_2=M_3/U(1)$ is in general 
an orbifold Riemann surface. This construction of course includes all the 
spherical three-manifolds in section~\ref{secondaryHopfgeometry},
for which $M_3\cong S^3/\Gamma_{ADE}$ and 
$\Sigma_2$ has genus $g=0$, but it also includes many 
other Seifert three-manifolds. For example, taking 
weights $(w_1,w_2,w_3)=(1,1,1)$ and $f$ to have degree $d$, then $M_3$ is the 
total space of a circle bundle over a Riemann surface
$\Sigma_2$ of genus $g=(d-1)(d-2)/2$.
 
Such homogeneous hypersurface 
singularities are Gorenstein canonical singularities, meaning they 
admit a global holomorphic $(2,0)$-form $\Omega_0$, defined on the 
complement of the isolated singularity at $Z_1=Z_2=Z_3=0$. 
With respect to the $I_+$ complex structure, so that we identify $\Omega_0=\Omega_+$, 
we may then write
\bea
\Omega_0 &=& \kappa\, \diff z\wedge \diff w~,
\eea
where $z$ and $w$ are the local coordinates defined by supersymmetry 
on $\R\times M_3$, defined in section \ref{sec:background},
and $\kappa=\kappa(z,w)$ is a local holomorphic function. The argument 
in section \ref{sec:unpairedprimary} then generalizes to give that the unpaired $B$ modes 
that contribute to the supersymmetric Casimir energy are 
\bea
B &  = &  \left|\frac{\mathcal{P}_+}{\Omega_+}\right|^{\qq/2} \mathscr{F}~,
\eea
where $|\mathcal{P}_+/\Omega_+|=\Omega^3 c/|\kappa|$ is a real, globally defined, nowhere zero function on $X\setminus\{o\}$
, and 
$\mathscr{F}$ is a holomorphic function on $X$. This follows since 
$\mathcal{P}_+$ and $\Omega_+$ are both globally defined, and being both $(2,0)$-forms are necessarily proportional. 
The holomorphic functions $\mathscr{F}$ on $X$ are  spanned by 
monomials ${Z}_1^{n_1}{Z}_2^{n_2}{Z}_3^{n_3}$, 
where $n_i\in\Z_{\geq 0}$, modulo the ideal generated by the defining polynomial $f$. 
The index-character that counts such holomorphic functions according  to their weights under $\Nlambda\in\C^*$  is
\bea\label{LX}
 \Lef(\bar{\partial},q,X) &=& \frac{1-q^d}{(1-q^{w_1})(1-q^{w_2})(1-q^{w_3})}~.
\eea

The $\phi$ modes work similarly, with respect to the second complex structure $I_-$. 
This may be defined globally in this setting as follows.  
The singularity $X$ may be viewed as a complex cone 
over the  orbifold Riemann surface $\Sigma_2=M_3/U(1)$. 
Here $\R\times M_3$ may be identified with a (orbifold) $\C^*$ fibration 
over $\Sigma_2$, with the isolated singularity arising by 
contracting the whole space to a point. In terms 
of the coordinates defined by supersymmetry, the 
$\C^*$ action is generated by the complex vector field $K$. 
The $I_-$ complex structure is then obtained by 
reversing the sign of the complex structure on the base $\Sigma_2$, while 
keeping that of the $\C^*$ fibre. This leads 
to the same complex manifold, although of course the map 
between the two copies is not holomorphic. As for the 
primary Hopf surfaces in section \ref{sec:primary}, the unpaired 
$\phi$ modes then give an identical contribution to the $B$ modes above.

It follows that the  relevant character is
\bea
\Lef(q,\lambda,X) &\equiv & q^{(-d+\sum_{i=1}^3w_i)/2}\lambda \cdot \Lef(\bar{\partial},q,X)~,
\eea
where $\Lef(\bar{\partial},q,X)$ is the index-character (\ref{LX}). 
Here the 
power of $q$ is precisely $\frac{1}{2}$ the charge of the holomorphic $(2,0)$-form (arising 
as usual since $A$ is a connection on $\mathcal{K}_+^{-1/2}$),
and $q\in \C^*$ is the generator 
of the $\C^*$ action. 
 The supersymmetric Casimir energy in this case is obtained as usual by setting $q=\ex^{t|b|}$,  $\lambda=\ex^{-tu}$, and extracting the coefficient of $-t$ 
 in a Laurent series about $t=0$. 
A simple calculation shows that this 
  leads to the supersymmetric Casimir energy
\bea\label{Esusyhyper}
\Esusym &=&  \frac{4du^3-(w_1^2+w_2^2+w_3^2-d^2)db^2u}{24b^2w_1w_2w_3}~.
\eea
Here $u=(r-1)\gamma$ for a matter multiplet of R-charge $r$, where now 
$1/2$ the Reeb charge of the holomorphic $(2,0)$ form is
$\gamma=(-d+\sum_{i=1}^3w_i)|b|/2$.
For example, the Lens space case $L(p,1)$ in sections \ref{sec:lens}, \ref{sec:fp}
is $w_1=2$, $w_2=w_3=p$,  $d=2p$ (with $|b_1|=|b_2|=|b|$), while the Poincar\'e Hopf surface in 
section \ref{sec:Poincare} is $w_1=10$, $w_2=6$, $w_3=15$, $d=30$.
We stress again that (\ref{Esusyhyper}) gives the matter contribution to 
the supersymmetric Casimir energy in the topological sector with trivial flat gauge connection. 
For non-trivial flat connections one would instead need to compute the index-character 
of the relevant (flat) matter bundle.

\subsection{Full supersymmetric Casimir energy}

As in much of the previous literature, in this paper we have focused attention on the contribution 
of a matter multiplet to the supersymmetric Casimir energy. However, 
we expect that the vector multiplet contribution will also arrange into short multiplets, and will similarly be  related to 
(twisted) holomorphic functions. At least for primary Hopf surfaces, and secondary Hopf surfaces with $M_3=L(p,1)$, 
previous results in the literature imply that the contribution of a vector multiplet to the 
supersymmetric Casimir energy is (formally) obtained from the contribution of a matter multiplet 
by (i) setting the R-charge $r=0$ (since the dynamical gauge field has zero R-charge), (ii) 
replacing weights $\rho$ by roots $\alpha$ of the gauge group $G$, and finally (iii) reversing the overall
sign. In this subsection we will simply \emph{conjecture} this is true more generally, at least 
in the sector with trivial flat connection on which we focus.

Given this conjecture, it is straightforward to combine the matter multiplet result (\ref{Esusyhyper}) 
for a general homogeneous hypersurface singularity with the vector multiplet result, and sum 
over relevant weights/roots. Remarkably, we find the following simple formula for the 
total supersymmetric Casimir energy
\bea\label{Esusytotal}
\Esusy &=& \frac{2|b|}{27}\frac{d c_1^3}{w_1w_2w_3}(3c-2a) + \frac{|b|}{3}\frac{dc_1}{w_1w_2w_3}(c_1^2-c_2)(a-c)~. 
\eea
Here we have defined
\bea
c_1 &\equiv & -d+\sum_{i=1}^3w_i~, \qquad c_2 \ \equiv \ -d^2 + \sum_{i=1}^3 w_i^2~,
\eea
which depend on the weights $(w_1,w_2,w_3)$ and degree $d$ of the hypersurface singularity, while 
$a$ and $c$ denote the usual trace anomaly coefficients, 
\bea
a &=& \frac{3}{32}(3 \mathrm{Tr}\, R^3 - \mathrm{Tr}\, R) \ = \ \frac{3}{32}\Big[2|G| + \sum_\rho \left(3(r_\rho-1)^3 - (r_\rho-1)\right)|\mathcal{R}_\rho|\Big]~,\\
c &=& \frac{1}{32}(9 \mathrm{Tr}\, R^3 - 5\mathrm{Tr}\, R) \ = \ \frac{1}{32}\Big[4|G| + \sum_\rho \left(9(r_\rho-1)^3 - 5(r_\rho-1)\right)|\mathcal{R}_\rho|\Big]~,\nn
\eea
with $R$ being the R-symmetry charge, and the trace running over all fermions.

By setting $(w_1,w_2,w_3)=(2,p,p)$, $d=2p$, which correspond to $A_{p-1}$ singularities with corresponding secondary Hopf 
surfaces $S^1\times L(p,1)$, one sees that (\ref{Esusytotal}) reduces to
\bea
\Esusy &=& \frac{16|b|}{27p}(3c-2a) + \frac{4|b|p}{3}(a-c)~.
\eea
This agrees with the $\beta\rightarrow\infty$ limit of the partition function in \cite{Nishioka:2014zpa}, and reproduces 
the original primary Hopf surface result of \cite{Assel:2014paa} when $p=1$.

One can make a number of interesting observations about the general formula (\ref{Esusytotal}). 
Firstly, it depends on the choice of supersymmetric gauge theory only via $a$ and $c$. 
Secondly, the coefficient of the term $(3a-2c)$ is related to the \emph{Sasakian volume} of $M_3$ via
\bea\label{Sasakianvol}
\vol(M_3) &=& \frac{d}{w_1w_2w_3}\cdot\frac{1}{|b|^2}\cdot \vol(S^3)~.
\eea
Here $\vol(S^3)=2\pi^2$ is the volume of the standard round metric on the unit sphere, 
and the Reeb vector is normalized as $\xi=|b|\zeta$, where $\zeta$ generates the 
canonical $U(1)\subset\C^*$ action on the hypersurface singularity.
$M_3$ is the link of this singularity, and 
any compatible Sasakian metric on $M_3$ has volume given by (\ref{Sasakianvol}), 
as follows from the general formula in \cite{Gauntlett:2006vf}.  The metric on $M_3$ 
is not in general Sasakian, but the point is that $M_3$ is equipped in general with an 
(almost) contact one-form $\eta=\diff\psi+a$. The corresponding contact volume $\frac{1}{2}\int_{M_3}\eta\wedge \diff\eta$ 
depends only on the Reeb vector, and thus agrees with the Sasakian volume. 
We shall briefly comment further on this in the discussion section. We also note that 
in (\ref{Esusytotal}) 
$c_1=-d+\sum_{i=1}^3w_i$ is the first Chern class (number) of the (orbifold)
anti-canonical  bundle of the orbifold Riemann surface $\Sigma_2=M_3/U(1)$ (more precisely, global
sections of  $K_{\Sigma^{\mathrm{orb}}}^{-1}$ are given by 
weighted homogeneous polynomials of degree $c_1$).
Thirdly, we have suggestively denoted $c_2=-d^2+\sum_{i=1}^3w_i^2$.
Of course this is not supposed to suggest the second Chern class/number of a line bundle, 
which is zero, but rather is a quadratic invariant of the singularity that takes a similar form to $c_1$. 
It would be interesting to understand the geometric interpretation of the second 
term, proportional to $(a-c)$, in (\ref{Esusytotal}). 

\section{Discussion}
\label{discuss:sec}

In this paper we have shown that the supersymmetric Casimir energy $\Esusy$ of four-dimensional ${\cal N}=1$ field theories defined on $S^1\times M_3$ is 
computed by a limit of the index-character counting holomorphic functions on (or more generally holomorphic sections over) the space $\R\times M_3$. In particular, the latter is equipped with 
an  ambi-Hermitian structure, and 
the short multiplets contributing to the supersymmetric Casimir energy  are in one-to-one correspondence with (twisted) holomorphic functions, with respect to either complex structure. 
As examples of Seifert three-manifolds $M_3$ we considered $S^3$, as well as the links $S^3/\Gamma_{ADE}$ of ADE hypersurface singularities in $\C^3$. For $M_3\cong S^3$ our analysis
explains the relation of the supersymmetric Casimir energy to the anomaly polynomial, pointed out in   \cite{Bobev:2015kza}. In the case of $M_3\cong L(p,1)$ we obtained 
formulas that may independently be derived using the 
path integral results of  \cite{Nishioka:2014zpa}; while,  to our knowledge,  the formulas for the $D$ and $E$ singularities have not appeared before. 
We have also presented a formula (\ref{Esusytotal}) for the 
supersymmetric Casimir energy when $M_3$ is the link 
of a general homogeneous hypersurface singularity, in the trivial flat connection sector, and assuming a conjecture 
for the vector multiplet contribution.

Our analysis can be extended in various directions. The localization results of
\cite{Assel:2014paa,Nishioka:2014zpa} strongly  suggest that in the supersymmetric quantum mechanics the 
contributions of the vector multiplet will also also arrange into short multiplets.  
One should show explicitly that these are indeed related to (twisted) holomorphic functions, and therefore ultimately to the index-character we have studied
(and in particular hence prove (\ref{Esusytotal})). 
In this paper we have explained how to incorporate the contributions of discrete flat connections on $M_3$, considering $M_3\cong L(p,1)$ as concrete example. 
It may be interesting to work out more examples. Moreover, here we have not addressed the role of continuous flat connections arising when $\pi_1(M_3)$ is infinite. 
Ultimately, the complete supersymmetric Casimir energy of a theory should be obtained by appropriately minimizing over the set of all flat connections, and it would be nice
to see whether this quantity may be used as a new test of dualities between different field theories and/or geometries. 

Using the formulas presented 
 in Appendix \ref{letters:sec} one can also easily obtain new supersymmetric indices for 
  theories defined on $S^1\times M_3$, where $M_3$ is the Seifert link of the $D$  and $E$ type hypersurface singularities. It would be interesting to explore their properties, as they involve a generalization of the elliptic gamma function appearing for $M_3\cong S^3$ \cite{Assel:2014paa} and $M_3\cong L(p,1)$ \cite{Benini:2011nc,Razamat:2013opa,Nishioka:2014zpa}. 

We close our discussion by recalling that it is  not clear how to reproduce the supersymmetric Casimir energy with a holographic computation in a supergravity solution, even for $M_3=S^3_{\mathrm{round}}$. See for example \cite{Cassani:2014zwa,Cassani:2015upa}  for some attempts and further discussion. Let us point out that the formula (\ref{Esusytotal}) shows that in the large $N$ limit the supersymmetric Casimir energy (in the trivial flat connection sector) is proportional to $N^2\cdot \mathrm{vol}(M_3)$.  We expect that it should be possible to reproduce this result from a dual holographic computation, and indeed we will report on this in \cite{Genolini:2016sxe}.

\subsection*{Acknowledgments}
\noindent
D.~M.~is  supported by the ERC  Starting Grant N. 304806, ``The Gauge/Gravity Duality and Geometry in String Theory''. 
J.~F.~S.~ was supported by the Royal Society in the early stages of this work. We thank Benjamin Assel for useful comments. 


\appendix 

\section{Supersymmertic index from the character}
\label{letters:sec}

In this appendix we  return to the supersymmetric index ${\cal I}$ 
\cite{Romelsberger:2005eg}, clarifying its relation to
 the index-character, that is the main subject of this paper. 

\subsection{Primary Hopf surfaces}
\label{lettersprimary:sec}

We begin with  the case $M_3\cong S^3$ and consider the modifications needed for the extension to more general $M_3$ in the next subsection. 
Following  \cite{Romelsberger:2005eg}, we can  work on  $M_4 = \R\times S^3_\mathrm{round}$,
 with the complex structure parameters of the Hopf surfaces emerging as 
fugacities associated to  two commuting global symmetries \cite{Closset:2013vra,Assel:2014paa}.  
The supersymmetric index may be defined
quite generally for any theory that admits the superalgebra (\ref{model}), in terms of a trace over states in the Hilbert space, as 
\bea
{\cal I} (x) & = & \mathrm{Tr}(-1)^F\, x^{\Sigma}~,
\eea
where $F$ is the fermion number. A standard argument then shows that the net contribution to the 
trace arises from states obeying $\Deltaqu \equiv  H_\mathrm{susy}-\Sigma = 0$. As this quantity does not depend on continuous parameters it can be computed in the 
free theory, where it takes the form of a plethystic exponential 
\bea
{\cal I}(x)  & = & \mathrm{Pexp} \left(f(x)\right)\, \equiv \, \mathrm{exp}\left(\sum_{k=1}^\infty \frac{1}{k} f(x^k) \right)  ~.
\eea
Physically, this is the grand-canonical partition function written in terms of the \emph{single particle} partition function $f(x)$, counting single particle states (annihilated by $\Deltaqu$)
of the 
free theory. In practice, the operator $\Sigma$ appearing in the superalgebra is given by 
$\Sigma =  -  (2J_3^L + R)$, where $R$ is the R-symmetry and $J_3^L$ is the angular momentum associated to rotations in $U(1)\subset SU(2)_L\subset SU(2)_L\times SU(2)_R$. One can introduce a second fugacity $y$ conjugated to the angular momentum $J_3^R$ associated to rotations in  $U(1)\subset SU(2)_R\subset SU(2)_L\times SU(2)_R$.
After changing variables\footnote{In this section we will denote  $p_1,p_2$ the variables in which the index is written naturally in terms of elliptic gamma functions. We will later 
make contact with the variables $q_1,q_2$ used in the previous sections.},  
setting $p_1=x y$ and $p_2= x/y$, the single particle index for a chiral multiplet is given by  \cite{Romelsberger:2007ec}
\bea
f^\mathrm{matter} (p_1,p_2)  & =  &\frac{(p_1p_2)^{\frac{r}{2}  } -  (p_1p_2)^{\frac{2-r}{2}}  }{(1-p_1)(1-p_2)}~, 
\label{deffsingle}
\eea
and the contribution of a chiral multiplet to the supersymmetric index then reads
\bea
{\cal I}^\mathrm{matter}  (p_1,p_2) = \prod_{n_1,n_2\geq 0}^\infty \frac{1- (p_1p_2)^{-r/2} p_1^{n_1+1}p_2^{n_2+1}}{1- (p_1p_2)^{r/2} p_1^{n_1}p_2^{n_2}}  
\ = \ \Gamma ((p_1p_2)^{r/2};p_1,p_2)~,\eea
where $\Gamma (z;p_1,p_2)$ is the elliptic gamma function.

It was noticed in \cite{Kim:2012ava,Assel:2014paa}  that the supersymmetric Casimir energy can be extracted from the single particle index by setting 
 $p_1=\me^{t|b_1|}$, $p_2=\me^{t|b_2|}$, and taking the finite 
part of the limit
\bea
E_\mathrm{susy} (|b_1|,|b_2|) &  = &  \frac{1}{2} \lim_{t\to 0}   \frac{\dd }{\dd t } f(p_1,p_2)~.
\label{oddprescription}
\eea
Below we will clarify the reason why this limit reproduces the supersymmetric Casimir energy by relating  $f^\mathrm{matter}  (p_1,p_2)$ to the index-character counting holomorphic functions. 

For the computation of   $f^\mathrm{matter}  (p_1,p_2)$   we can use  the ingredients worked out in  \cite{Gerchkovitz:2013zra,Lorenzen:2014pna}. In particular, the expressions for the operators 
$H_\mathrm{susy}, R, J_3^L, J_3^R$ can be found in these references\footnote{We use the notation of \cite{Lorenzen:2014pna}. For simplicity, and to make contact with 
 \cite{Romelsberger:2007ec}, we are setting the parameters $\kappa$, $\epsilon$ in \cite{Lorenzen:2014pna} to $\kappa=-1$, $\epsilon=1$.}, written 
 in terms of bosonic  and fermionic  oscillators. For example,  writing
$\Deltaqu = \Deltaqu_\mathrm{bos} +  \Deltaqu_\mathrm{fer}$, we have
\be
\Deltaqu_\mathrm{bos}   = \frac{1}{2} \sum_{\ell=0}^\infty\sum_{ m,n =-\frac{\ell}{2}}^{\frac{\ell}{2}} \Deltaqu^a_{\ell  m }\left( a_{\ell mn } a_{\ell mn }^\dagger +a_{\ell mn }^\dagger a_{\ell mn } \right) + \Deltaqu^b_{\ell m } \left(b_{\ell mn } b_{\ell mn }^\dagger  + b^\dagger_{\ell mn } b_{\ell mn } \right)  \label{scalarDelta}  ~ ,
\ee
and 
\bea
\Deltaqu_\mathrm{fer} & = & \frac12  \sum_{\ell=0}^\infty \sum_{n=-\frac{\ell}{2}}^{\frac{\ell}{2}} \sum_{m=-\frac{\ell}{2}-1}^{\frac{\ell}{2}} \Deltaqu^c_{\ell m} \left(  c_{\ell m n } c^\dagger_{\ell m n}-c_{\ell m n }^\dagger c_{\ell m n} \right) \nn\\
&& - \frac12 \sum_{\ell=1}^\infty \sum_{n=-\frac{\ell}{2}}^{\frac{\ell}{2}} \sum_{m=-\frac{\ell}{2}}^{\frac{\ell}{2}-1} \Deltaqu^d_{\ell m}\left(  d_{\ell m n} d^\dagger_{\ell m n} - d_{\ell m n }^\dagger d_{\ell m n} \right)  
\label{fermDelta} ~,
\eea
with 
\bea
\Deltaqu^a_{\ell m}  & = &   \ell +  2+  2m ~ ,\qquad \quad ~\, ~\Deltaqu^b_{\ell m}   \ = \    \ell + 2m ~, \nonumber\\
 \Deltaqu^c_{\ell m}   & = &  -   \left( \ell +  2+  2m \right) ~, \qquad \Deltaqu^d_{\ell m}  \ = \    -\ell + 2m  ~,
 \eea 
and  similar expressions for the other operators. 
There are four types of  single particle states in the Fock space, namely 
$| a_{\ell, m, n} \rangle =  a_{\ell m n}^\dagger |0 \rangle $, $| b_{\ell, m, n} \rangle =  b_{\ell m n}^\dagger |0 \rangle $, $| c_{\ell, m, n} \rangle =  c_{\ell m n}^\dagger |0 \rangle $, 
and $| d_{\ell, m, n} \rangle =  d_{\ell m n}^\dagger |0 \rangle $. However,  the only zero-modes of $\Deltaqu$
are 
\bea
 | b_{\ell, -\frac{\ell}{2}, n}\rangle ~, \qquad \qquad  | c_{\ell, -\frac{\ell}{2}-1, n}\rangle ~, 
 \label{singlepmodes}
\eea
while there are no zero-modes of the $a$-type and $d$-type states.  These have   $m=-\frac{\ell}{2}$ and  $m=-\frac{\ell}{2}-1$, respectively, which are precisely the shortening conditions obeyed by the $\phi$ and $B$ modes, in the special case of the round three-sphere \cite{Assel:2015nca}. 
These two sets of modes are contributing non-trivially to (\ref{deffsingle}). Let us now show this explicitly. From the definition
\be
\label{fxy}
f^\mathrm{matter}  (x,y) \ = \ \mathrm{tr}(-1)^F  x^\Sigma y^{2 J_3^R} \ = \  f_\mathrm{bos} (x,y) - f_\mathrm{fer} (x,y)~, 
\ee
where here the trace is over the single particle states in (\ref{singlepmodes}), and we have
\bea
f_\mathrm{bos} (x,y) & =&  \sum_{\ell=0}^\infty      x^{r+\ell} \sum_{n=-\tfrac{\ell}{2}}^{\tfrac{\ell}{2}}    y^{2n} \ = \  \frac{x^r}{(1- x y )(1-\tfrac{x}{y})}~,\nonumber\\
f_\mathrm{fer} (x,y) &=  &\sum_{\ell=0}^\infty      x^{\ell -r+2} \sum_{n=-\tfrac{\ell}{2}}^{\tfrac{\ell}{2}}    y^{2n} \ =\  \frac{x^{2-r}}{(1- x y )(1-\tfrac{x}{y})}~.
\eea
To derive these we used\footnote{Here the operators are \emph{normal ordered}   \cite{Romelsberger:2007ec}.} 
\be
\Sigma \,  | b_{\ell, -\frac{\ell}{2}, n} \rangle  \ = \ (r+\ell)    | b_{\ell, -\frac{\ell}{2}, n} \rangle ~,\qquad 
\Sigma \,   | c_{\ell, -\frac{\ell}{2}-1, n}\rangle  \ =\   - (r- 2-\ell)   | c_{\ell, -\frac{\ell}{2}-1, n}\rangle~,
\ee
and 
\be
J_3^R \,  | b_{\ell, -\frac{\ell}{2}, n} \rangle \  = \ n    | b_{\ell, -\frac{\ell}{2}, n} \rangle ~,\qquad 
J_3^R  \,   | c_{\ell, -\frac{\ell}{2}-1, n}\rangle  \ = \   n  | c_{\ell, -\frac{\ell}{2}-1, n}\rangle~.
\ee
Notice that  the R-charge of the bosonic modes  $ | b_{\ell, -\frac{\ell}{2}, n}\rangle $  is $r$, while that of the fermionic modes 
$ | c_{\ell, -\frac{\ell}{2}-1, n}\rangle $ is $-(r-2)$. Thus  $f^\mathrm{matter}  (x,y)$ is counting the bosonic particles minus the fermionic anti-particles
\cite{Romelsberger:2007ec}. 

In order to make contact with the main part of the paper, one can see that upon making the identifications\footnote{The need for this change of variables originates from 
our definition of the complex structures. See footnote \ref{fucsigns}. This is of course just a convention.} $p_1=q_1^{-1}$, $p_2=q_2^{-1}$, 
the first term in (\ref{deffsingle}) is precisely the character $\Lef(\bar{\partial}_{\mathcal{K}_-^{r/2}},(q_1,q_2))$ in (\ref{phich}),
counting $\phi$ modes. On the other hand,  the second term is equal to  the character $\Lef(\partial_{\mathcal{K}_+^{-(r-2)/2}},(q_1,q_2))$, 
namely it can be identified with the character counting $\tilde B$ modes. Notice that 
\bea
   \Lef(\partial_{\mathcal{K}_+^{-(r-2)/2}},(q_1,q_2))& = &    \Lef(\bar{\partial}_{\mathcal{K}_-^{r/2}},(q_1^{-1},q_2^{-1})) ~.
\eea
On taking the limit (\ref{oddprescription}), the opposite signs  in front of the fermionic part and in its exponent  cancel each other, effectively giving 
the \emph{same} result as the  limit of the character, or Dirac character, that we considered before.   

\subsection{Secondary Hopf surfaces}
\label{letterssecondary:sec}

Let us now discuss  secondary Hopf surfaces $M_4=S^1\times M_3$,  starting with the case that  the fundamental group of $M_3$ is
 $\Gamma\cong \Z_p$. Thus $M_3=L(p,1)$ is a Lens space. The supersymmetric index in this case was studied in  \cite{Benini:2011nc,
Razamat:2013opa,Nishioka:2014zpa}. We can work on the space with a round metric on $S^1\times S^3/\Z_p$ and obtain the modes 
by \emph{projecting} from those on the covering space $S^1\times S^3$. In the absence of a flat connection 
the modes on  $L(p,1)$  are  precisely  the $\Z_p$-invariant modes on $S^3$. For example, for the scalar field $\phi$, these are given by the 
$S^3$ hyperspherical harmonics $Y_\ell^{mn}$ satisfying  $2n \equiv 0 ~\mathrm{mod} ~p$.
More generally, in the presence of a flat connection with first Chern class $c_1(L)$, the modes that descend to the Lens space from $S^3$ obey the condition
 \cite{Benini:2011nc,Alday:2012au}
\be
2n \ \equiv \ c_1(L) ~\mathrm{mod} ~p~.
\label{genproj}
\ee

Since the flat connection can be removed locally by a gauge transformation, the eigenvalues of the operators $H_\mathrm{susy}, R, J_3^L, J_3^R$ are unchanged.  One can then compute the generating function by restricting the sums in (\ref{fxy}) to the single particle states annihilated by $\Deltaqu$ of the previous subsection, and further obeying the projection (\ref{genproj}), with $c_1(L)= \rho (\mathbf{m})=\holonomy$. Accordingly, the bosonic part is then given by 
\bea
f_\mathrm{bos} (x,y) & =&  x^r  \sum_{\ell=0}^\infty      x^{\ell} \sum_{n \in P }     y^{2n} ~,
\label{sumlensproj}
\eea
where $P = \{ n  \in \{-\frac{\ell}{2},\dots,\frac{\ell}{2} \}  :  ~ 2n \equiv \nu  ~\mathrm{mod} ~p \}$. The sums are then computed exactly 
as in section      \ref{sec:lens}, and  we have 
\bea
f_\mathrm{bos} (x,y) &=& x^r \frac{(xy)^\holonomy (1- x^{2(p-\holonomy)}) + (\tfrac{x}{y} )^{p-\holonomy} (1-x^{2\holonomy})} {(1- x^2) (1 - (xy)^{p}) (1- (\tfrac{x}{y})^p)}~.
\eea
Expressing  this in terms of the variables $p_1=x y$ and $p_2= x/y$, we obtain
\bea
f_\mathrm{bos}^{p,\holonomy} (p_1,p_2)  & =  &    (p_1p_2)^{\frac{r}{2}}  \, \Lef(\bar{\partial}_L,(p_1,p_2),\C^2/\Z_p)~.
\label{fullflens}
\eea

For the fermions in the complex conjugate multiplet, the projection condition  has to be modified as  \cite{Razamat:2013opa}
 \be
2n \equiv - \holonomy ~\mathrm{mod} ~p~.
\label{antigenproj}
\ee
This effectively swaps $n_1$ and $n_2$, or equivalently, $p_1$ and $p_2$. Therefore, the index counting antifermions is given by 
\bea
f_\mathrm{fer}^{p,\holonomy}(p_1,p_2)  & =  &  (p_1p_2)^{\frac{2-r}{2}}  \, \Lef(\bar{\partial}_L,(p_2,p_1),\C^2/\Z_p)~,
\eea
Again, it can be checked explicitly that 
$f_\mathrm{fer}^{p,\holonomy} (p_1,p_2)  =  f_\mathrm{bos}^{p,\holonomy} (p_1^{-1},p_2^{-1})$, showing the character contributing to the fermions is counting \emph{anti}-holomorphic sections, as opposed to the bosonic character, which counts   holomorphic sections. 
Of course, the result of the limit (\ref{oddprescription})  reproduces precisely the supersymmetric Casimir energy in (\ref{esusylens}). 

In order to compute the supersymmetric index using the plethystic exponential, it is convenient to write the  twisted Lens space character as
\bea
 \Lef(\bar{\partial}_L,(p_1,p_2),\C^2/\Z_p) 
& = &   \frac{p_1^\holonomy}{(1-p_1p_2)(1-p_1^p)} +   \frac{p_2^{p-\holonomy}}{(1-p_1p_2)(1-p_2^p)} ~.
\label{factorclens}
\eea
Using this, it is immediate to obtain the index in the factorised form \cite{Razamat:2013opa}, namely
\bea
{\cal I}_{\, p,\holonomy}^\mathrm{matter}  (p_1,p_2)  \ = \  \Gamma ( (p_1p_2)^{\frac{r}{2}}p_2^{p-\holonomy};  p_2^p,  p_1p_2)\,  \Gamma ( (p_1p_2)^{\frac{r}{2}}p_1^{\holonomy}; p_1^p,  p_1p_2)~,
\label{factorized}
\eea
where notice that this does not contain any Casimir energy contribution. 
   
  The reasoning that led to the expression of the single particle index above should be  valid more generally for a theory defined on  
$M_4=\R \times M_3$ (where $\pi_1(M_3)$ is finite), with a fixed flat connection  in a sector $\iflat\in\Mflat$.
In particular, we expect that this is always given by 
\bea
f^\mathrm{matter}  (p_1,p_2)  \ =  \  (p_1p_2)^{\frac{r}{2}}  \, \Lef(\bar{\partial}_\alpha,(p_1,p_2),M_4) -  (p_1p_2)^{-\frac{r}{2}}  \, \Lef(\bar{\partial}_\alpha,(p_1^{-1},p_2^{-1}),M_4) ~.
\label{fullcorrect}
\eea
 However, we will not pursue this direction further here. To illustrate our prescription,  below
 we will derive expressions 
for the (chiral multiplet contribution to the) supersymmetric index in the class of homogeneous hypersurface singularities, 
in the sector without flat connection.

As before, to evaluate the bosonic single letter partition function, we can start from the theory on $\R\times S^3$, and evaluate the sums as in (\ref{sumlensproj}) by projecting out the modes not invariant under $\Gamma \subset SU(2)$. This is equivalent to counting 
holomorphic functions on $\C^2$ that are invariant under $\Gamma$. For $\Gamma=\Z_p$ this is of course the case of the Lens space, yielding (\ref{factorclens}). 
Let us then discuss the remaining $D$ and $E$ singularities.  Implementing the projection on the modes, we find 
\bea\label{fxDE}
f_\mathrm{bos}^{DE} (x)  &=& x^r \frac{1-x^{2d}}{(1-x^{2w_1})(1-x^{2w_2})(1-x^{2w_3})}~,
\eea
where the weights and degrees of the singularities can be read off from the defining equations given in (\ref{hsequations}). For example, for the
$E_8$ singularity, corresponding to the Poincar\'e Hopf surface, the (minimal) set of weights is $(w_1,w_2,w_3)=(10,6,15)$, with 
 degree  $d=30$. For the $D_{p+1}$ series the weights are $(w_1,w_2,w_3)=(2,p-1,p)$ and the degree is $d=2p$.
Notice that in all cases the series expansion of (\ref{fxDE}) does not contain odd powers of $x$. This is because for $\Gamma=\Gamma_D$ and 
$\Gamma=\Gamma_E$, $\Gamma \supset \Z_2$, where this acts as $\Z_2: (z_1,z_2)\to - (z_1,z_2)$.

Changing variable setting $x=q^{1/2}$, we indeed find that 
\bea\label{fqDE}
f_\mathrm{bos}^{DE} (q)  &=& q^{r/2} \frac{1-q^{d}}{(1-q^{w_1})(1-q^{w_2})(1-q^{w_3})} \ = \   q^{r/2}  \Lef(\bar{\partial},q,\C^2/\Gamma) ~.
\eea
Moreover, using that 
\bea
w_1 + w_2  + w_3 -d & = & 1~,
\label{good}
\eea
 we compute 
\bea\label{fferqDE}
f_\mathrm{fer}^{DE} (q)  &=&   f_\mathrm{bos}^{DE} (q^{-1}) \ = \    q^{(2-r)/2}  \Lef(\bar{\partial},q,\C^2/\Gamma) ~.
\eea
Thus the single particle index for the chiral multiplet reads
\bea\label{fchiDE}
f^\mathrm{matter}_{DE} (q)  &=&  \frac{(q^{r/2}- q^{(2-r)/2})  (1-q^{d})}{(1-q^{w_1})(1-q^{w_2})(1-q^{w_3})}   ~,
\eea
and taking the plethystic exponential it results in the following  \emph{triple} infinite products
\bea
{\cal I}^\mathrm{matter}_{DE} (q)  =  \frac{\prod_{n_1,n_2,n_3\geq 0}^\infty \left(1 - q^{1-r/2}q^{n_1w_1+n_2w_2+n_3w_3} \right) \left(1 - q^{r/2+d}q^{n_1w_1+n_2w_2+n_3w_3} \right)}{\prod_{n_1,n_2,n_3\geq 0}^\infty \left(1 - q^{r/2}q^{n_1w_1+n_2w_2+n_3w_3} \right) \left(1 - q^{1-r/2+d}q^{n_1w_1+n_2w_2+n_3w_3} \right)}~.
\label{generalXindex}
\eea
Notice that this cannot be expressed in term of the ordinary elliptic gamma functions. 
However, interestingly, using the condition (\ref{good}), valid for the $D$ and $E$ singularities, we find that this can be written as 
\bea
{\cal I}^\mathrm{matter}_{DE} (q) & = &  \frac{\Gamma ( q^{r/2+d};  q^{w_1},  q^{w_2}, q^{w_3})}{\Gamma ( q^{r/2} ;  q^{w_1},  q^{w_2}, q^{w_3}) }~,
\eea
where 
\bea
\Gamma  (z; q_1,q_2,q_3)  \ =  \prod_{n_1,n_2,n_3\geq 0}^\infty \!\! (1 - z^{-1} q_1^{n_1+1} q_2^{n_2+1} q_3^{n_3+1} ) \,  (1 - z q_1^{n_1} q_2^{n_2} q_3^{n_3} ) ~
\label{triplegamma}
\eea
is a generalization of the elliptic gamma function \cite{Spiridonov:2012de,nishizawa}.

\end{document}